\documentclass[a4paper,12pt]{amsart}
\usepackage{amssymb,amsmath,mathrsfs}
\usepackage[usenames,dvipsnames]{xcolor}
\usepackage{hyperref}
\usepackage{tensor}
\usepackage{graphicx}
\usepackage[left=1in,top=1in,right=1in,bottom=1in,headheight=0.8in,foot=0.5in]{geometry}
\usepackage[nodisplayskipstretch]{setspace} 
\usepackage{mdframed}
\usepackage{makecell}
\usepackage{tabularx}
\usepackage{bbm}
\usepackage{amsthm}
\newtheorem{definition}{Definition}
\usepackage{array}
\usepackage{booktabs}
\usepackage{geometry}
\usepackage[greek.polutoniko,english]{babel}
\usepackage{amsthm}

\newtheorem{proposition}{Proposition}

\hypersetup{
	colorlinks=true,         
	linkcolor=MidnightBlue,          
	citecolor=MidnightBlue,
	urlcolor=MidnightBlue            
 }
 \let\oldmarginpar\marginpar
\renewcommand\marginpar[1]{\oldmarginpar{\color{red}\raggedright\scriptsize #1}}
\let\oldmarginpar\marginpar
\renewcommand\marginpar[1]{\oldmarginpar{\color{red}\raggedright\scriptsize #1}}
\newcommand{\pb}[2]{\ensuremath{\lf\{#1,#2 \rt\}}}

\newcommand{\diby}[2]{\ensuremath{\frac{\partial #1}{\partial #2}}}

\newcommand{\comment}[1]{}
\newcommand{\bT}{\mathbf{T}}

\newcommand{\Abs}[1]{\ensuremath{\text{Abs}(#1)}}
\newcommand{\Rel}[1]{\ensuremath{\text{Rel}(#1)}}
\newcommand{\Inc}[1]{\ensuremath{\text{Inc}(#1)}}

\def\de {\ensuremath{\text{d}}}
\def\lf {\ensuremath{\left}}
\def\rt {\ensuremath{\right}}

\def\mS {\ensuremath{\mathcal S}}

\def\mT {\ensuremath{\mathcal T}}
\def\Lie { \ensuremath{\mathfrak {L}} }

\usepackage{natbib}
\setcitestyle{aysep={}} 

\title{\sc Nomic Structure and Reduction}
\date{Draft of \today.}

\usepackage{amsaddr}
\usepackage{float}

\author{Sean Gryb}
\address{\vspace{-0.8pc}University of Groningen}
\email{\href{s.b.gryb@rug.nl}{s.b.gryb@rug.nl}}

\author{Karim P. Y. Th\'ebault}
\address{\vspace{-0.8pc}University of Bristol}
\email{\href{mailto:karim.thebault@bristol.ac.uk}{karim.thebault@bristol.ac.uk}}

\makeatletter
\let\uppercasenonmath\@gobble

\begin{document}
\setstretch{1.1}
\begin{abstract}
The canonical formulation of physical theories with \textit{irregular nomic structure} is as constrained Hamiltonian theories within which ill-posedness of the equations of motion is connected to a \textit{pernicious} form of surplus representational capacity. Such theories can be converted into theories with regular nomic structure and a well-posed initial value problem via the process of symplectic reduction. We analyse, synthesise, and contrast different approaches to the presentation and analysis of constrained Hamiltonian theories, drawing upon recent work on formalisation of nomic structure on model spaces \citep{gryb:2024} and comparisons of theoretical structure and representational capacity via category theory \citep{bradley:2020,bradley:2025}. We suggest that the case of irregular nomic structure is most naturally suited to a category theoretic presentation in which \textit{state spaces are arrows} and \textit{symplectic reduction is arrow composition} \citep{landsman:2005}. Under this approach one obtains the natural results that theories with isomorphic state spaces are equivalent and theories whose reduced state spaces are isomorphic are equivalent at the level of the regular representations of their nomic structure. This analysis provides a suitable foundation for the case of quantization of theories with irregular nomic structure, which will be in a companion paper.
\end{abstract}

\maketitle

\tableofcontents

\setstretch{1.3}

\section{Introduction}

The nomic structure of a physical theory is the structure which encodes the laws of that theory. In the context of representations of physical theories as model space we find nomic structure playing two fundamental and distinct roles. The first, and most basic, is as a partitioning function whereby the nomic structure indicates which models are \textit{merely kinematically possible} and which models are \textit{dynamically possible}. The second, equally important and yet often neglected, is a projection function whereby the nomic structure indicates which of the set of dynamically possible models can be understood as distinct and which equivalent. Significantly, the full specification of a dynamical equivalence principle will depend upon the representational context and will thus require physical as well as purely formal inputs cf.  
\cite{Belot:2013,Wallace:2019}. However, crucially there are purely formal conditions under which we are \textit{required} to project out distinctions between at least some models according to some dynamical equivalence principle, else the nomic structure will not provide well-posed dynamical equations. In such cases we say that the nomic structure is \textit{irregular}. In cases in which the relevant formal conditions do not obtain we say that the nomic structure is \textit{regular}.

Significantly, the philosophical literature on theoretical structure, representation, and equivalence is largely confined to the regular case. In such contexts, an increasingly sophisticated account has been progressively developed in which the tools of category theory have been applied towards the diagnosis of inter-theory relations.\footnote{In what follows we will primarily follow the account provided in \cite{bradley:2020}. See also \cite{Halvorson2012,Halvorson2016,Weatherall2019a,Weatherall2019b,nguyen:2020,Barrett2020a,Barrett2020b,Dewar2022,feintzeig:2025,bradley:2025} and further canonical sources cited therein.} The standard approach is to treat theories as categories with models as the objects and the morphisms are the isomorphisms of the theory. Conditions on functors between the theories, understood as categories, can then be applied towards the formal analysis of claims regarding, for example, theoretical equivalence or the existence of surplus structure. In a recent article \cite{bradley:2025} has broken new ground by extending the application of category theory as a tool to the analysis of the structure of a theory with irregular nomic structure. The particular virtues of her account are that it allows for both a clear articulation of the sense in which representation of theories with irregular nomic structure as constrained Hamiltonian theories include \textit{surplus structure}  and provides tools to understand the \textit{symplectic reduction} process for converting irregular to regular nomic structure in terms of pair of functors between categories with models as the objects. The present article is intended as a complement to this important work both by supporting Bradley's key conclusion regarding surplus structure via additional physical argumentation and in offering an alternative means of category theoretic presentation.  

In the first regard, our particular goal is to present a number of formal and physical arguments that demonstrate the necessity of implementing dynamical equivalence principles in the context of irregular theories supporting. Our central concern will be the formal analysis of the initial value problem of classical mechanical theories and we will provide an overview of material relating to problems of ill-posedness in the context of the Euler-Lagrange equations, contained Hamiltonian theories, and first-order geometric velocity-phase space formalism. Our analysis will both provide a synthetic analysis of famous results due to Noether and Dirac, demonstrating the limitations of these approaches, and, crucially, articulate a more general and physically transparent set of diagnostic criteria that allow for the isolation of initial value constraints that indicate the existence of dynamical redundancy that generates ill-posedness.\footnote{This work builds on earlier analysis in \cite{gryb:2024} and is complemented by a more extensive treatment found in \cite{Gryb:2026}.}

In the second regard, our goal is to consider an alternative category theoretic presentation. In particular, we argue that the case of irregular nomic structure is most naturally suited to a category theoretic presentation in which \textit{state spaces are arrows} and \textit{symplectic reduction is arrow composition} \citep{landsman:2005}. Under this approach one obtains the natural results that theories with isomorphic state spaces are equivalent and theories whose reduced state spaces are isomorphic are equivalent at the level of the regular representations of their nomic structure. Furthermore, this analysis provides a suitable foundation for the case of quantization of theories with irregular nomic structure, which will be in a companion paper.

Our analysis will proceed according to the following plan. Section \ref{TheoreticalStructure} is devoted to the abstract characterisation of theoretical structure and provides a constructive synthesis of a number of lines of existing work. We start by articulating the core ideas being the notions of nomic structure, dynamical distinctness, and irregularity following the account of \cite{gryb:2024} and then proceed to present the core features of the standard category theory  approach to theoretical structure following \cite{bradley:2020}, before finally articulating an account of symmetry and structure that is implied by the intersection of these two approaches. 

Section \ref{Irregularity} then provides an analysis of irregular nomic structure starting from the Euler-Lagrange equations and proceeding through various articulations of the canonical analysis including the work of  \cite{bradley:2025} and the famous results of Noether and Dirac.  We provide a high-level overview of the novel results of \cite{gryb:2024} and \cite{Gryb:2026} which resolve ambiguities in earlier work and result in a physically transparent set of diagnostic criteria   for isolation of the dynamical redundancy that generates ill-posedness in terms of the existence of initial value constraints. 

Finally, in Section \ref{Reduction and Nomic Regularity} we consider various aspects formal and interpretative presentations  framework of symplectic reduction with a focus on its role in the conversion of an irregular to regular nomic structure. We first recap the result of \cite{bradley:2025} under which the quotenting aspect of symplectic reduction can be rendered as a particular type of functorial relation between categories with models as objects. We then introduce the formal machinery needed to understand the alternative approach due to \cite{landsman:2001,landsman:2005} which builds on earlier work by \cite{weinstein:1983}. In particular, we first introduce the crucial ideas of momentum maps and symplectic dual pairs. This then allows us to understand state spaces as arrows and symplectic reduction as arrow composition leading to the natural results mentioned above and, as set out in the short prospectus, a basis upon which one is able to articulate the formal tools and interpretational concepts needed to understand the quantization of theories with irregular nomic structure. 

\section{Theoretical Structure}
\label{TheoreticalStructure}

\subsection{Model Spaces}

The notions of \textit{constitutive structure} and \textit{nomic structure} of a theory are based upon an approach in which theories are presented in terms of model spaces. Following the framework introduced by \cite{gryb:2024} we will apply the following definitions: The \textit{constitutive structure} of a theory is the structure that one must assume in order to build the space, $K$, of \textit{kinematically possible models} (KPM) of that theory. That is, the space which represents the basic `pre-nomic' set of possibilities admitted by the theory. Each of these models can be thought of as something like a bare universe, stripped of laws and dynamics. Constitutive structures are often, although not aways, provided in terms of state space structures, used to characterize physical events, geometric or topological structures, used to characterize relations between the events, and matter structures, used to characterize particles or fields. Models with different \textit{tokens} of the same \textit{type} of constitutive matter and geometric structure are typically constitutively distinct KPMs of the same theory. Thus, for example, two- and three- body particle models can be understood as having the distinct tokens of a common Newtonian constitutive structure. 

The token-type distinction  allows us to distinguish between the following two cases. First we have structure common between all KPMs which share the same \textit{type} of structure. Such structure is \textit{constitutively fixed}. Constants of nature can typically be understood as an example of constitutively fixed structures of a theory. Second we have structure that is common between all KPMs which share the same \textit{token} of structure but which varies between at least two distinct tokens of the same type of constitutive structure. Such structure is \textit{contingently fixed}. Constants of motion can typically be understood as an example of contingently fixed structures of a theory. This distinction allows for differentiation between the modal status of two different types of structure. That is, structure that is the same in all KPMs of the same model type, and structure that can vary between kinematically possible models that instantiate different tokens of that type. It is worth noting that with regard to KPMs, the framework we are using takes a form similar to that used to characterize the model space of a physical theory by, for example, \cite{pooley:2017}. This approach is articulated in terms of a specification of the space of KPMs together with a set of solution-independent `fixed' (or `non-dynamical') fields common to all KPMs, and a set of `non-fixed' (or `dynamical')  fields not common between all KPMs. Constitutively fixed in our sense thus coincides with fixed in the sense of \cite{pooley:2017}.

Nomic structure is then the structure of a theory that represents the laws of that particular theory. Nomic structure has two important and distinct functions. The first function is to \textit{partition} the space of KPMs: that is, to tell us which models are dynamically possible and which are not dynamically possible.  We can represent this function partitioning of the space of KPMs into the proper sub-space of dynamically possible models, or DPMs, $D\subset K$. The second function is to provide us with an equivalence relation between DPMs. The equivalence relation function provides a methodology for determining which DPMs are dynamically distinct and which are dynamically identical. The relevant notion of distinctness and identity is here a nomic one rather than an ontic one; that is, a distinction based upon a difference that the laws pick out between two models and not a distinction that is necessarily equivalent to a strong metaphysical notion of distinctness and identity. 

\begin{figure}[htbp]     
\begin{center}
\includegraphics[height=4cm]{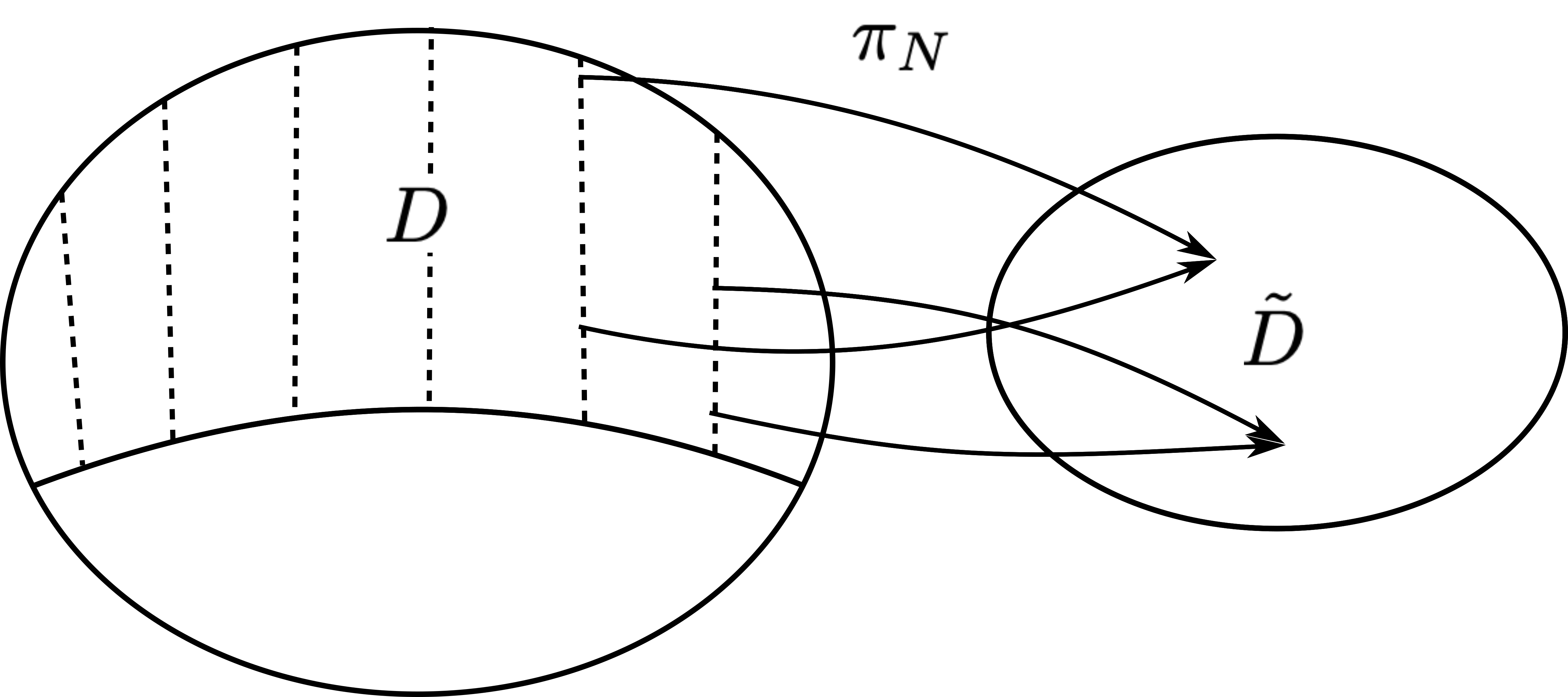}
\caption{Schematic representation of the nomic structure. $D$ is the partition of dynamically possible models. The dotted lines are `fibres' that represent dynamically equivalent models which the projection map, $\pi_N$, maps into single points in the space of distinct dynamically possible models, $\tilde D$.} \label{nomicstructure1}    
\end{center}
\end{figure}

We can represent the equivalence relation function of nomic structure in terms of a \textit{Projection Map}, $\pi_N$. We will provide a more formal definition of the map later but the important point for our purposes is that the map is such that it removes distinctions between models according to some dynamical equivalence principle. We can designate the space of equivalence classes $\tilde D$ of $D$ the space of Distinct Dynamically Possible Models (DDPMs); this allows us to consider the projection from the space of DPMs to the space of DDPMs: $\pi_N: D \to \tilde D$. The role of the nomic structure can be given a schematic representation as per Fig. \ref{nomicstructure1} where we have introduced the terminology of a `fibre' for the equivalence class of DPMs.

Significantly, the full specification of a dynamical equivalence principle will depend upon the representational context and will thus require physical as well as purely formal inputs cf.  
\cite{Belot:2013,Wallace:2019}. However, crucially, as discussed in \cite[5.3]{gryb:2024} and elaborated below, there are purely formal conditions under which we are \textit{required} to project out distinctions between at least some models according to some dynamical equivalence principle, else the nomic structure will not provide well-posed dynamical equations. In such cases of \textit{irregular nomic structure} the projection map \textit{cannot} be treated as the identity and at least some DPMs \textit{must} be treated as dynamically equivalent. It is \textit{mandatory} to apply a non-trivial criterion of dynamical equivalence precisely because the DPMs in question do not correspond to independently well-posed solutions of the equations of motion and implementing a non-trivial projection is a \textit{sine qua non} of a non-pathological specification of nomic structure. 

The contrast case is that of \textit{regular nomic structure}. In such theories the projection map \textit{may} be treated as the identity and we have that all DPMs \textit{may} be treated as distinct (i.e. $D = \tilde D, \pi_N: D \to  D$). Identification of dynamical equivalence requires physical or interpretative input. For example, in Newtonian particle mechanics there are good physical reasons coming from the variational principle to take DPMs related by rigid, time independent coordinate transformations to be dynamically equivalent and thus implement a non-trivial projection $\pi_N: D \to \tilde D$ where the fibres are ISO(3) transformations \cite[6.5]{gryb:2024}. There is, however, nothing formally inconsistent in taking such `Leibniz shifted' DPMs as dynamically distinct since they correspond to independently well-posed solutions of the equations of motion.

Crucially, the difference in the interpretation of regular and irregular theories maps onto a formal difference in both the canonical representation and the admissible quantization procedures relevant to regular and irregular classical theories of mechanics. It is thus much more than a \textit{merely philosophical} distinction. One main aim of this paper is to explicate the structure of theories with irregular nomic structure by application of ideas from the literature of category theory and theoretical structure. We review some of these ideas in the following sub-sections. Section \ref{Irregularity} is then devoted to considering the formal analysis of classical theories with irregular nomic structure represented as constrained Hamiltonian theories with a particular focus on the problem of ill-posed initial value problems.      
 
\subsection{Theories as Caterogies}

 Let us define a category $\mathcal{C}$ as a collection of \textit{objects} and a collection of \textit{morphisms} or \textit{arrows} (we will use the two terms interchangeably), each with a source and target object. The morphisms or arrows in question will typically be invertible and thus isomorphisms in the applications we will consider. We denote a morphism by $f : A \to B$ with the source $A$ and target $B$. The collection of morphisms with source $A$ and target $B$ in the category $\mathcal{C}$ is called the hom-set between $A$ and $B$ and is denoted $\mathrm{Hom}_\mathcal{C}(A,B)$. A category comes equipped with an operation of composition for morphisms, denoted $\circ$, which is total, associative and such that for each object $A$ there is a unique identity morphism $1_A$ whose composition with other morphisms leaves them unchanged. Thus we have a category $\mathcal{C}$, objects $(A,B)$ and morphisms $f,g \in \mathrm{Hom}_\mathcal{C}(A,B)$ such that:
$f : A \to B, g : B \to C, f \circ g = h : A \to C$. We can then define a \textit{functor} $F : \mathcal{C} \to \mathcal{D}$ between two categories $\mathcal{C}$ and $\mathcal{D}$ in terms of a pair of maps: one map between the objects of $\mathcal{C}$ and the objects of $\mathcal{D}$ and another map between the morphisms of $\mathcal{C}$ and the morphisms of $\mathcal{D}$. By definition a functor preserves sources and targets and arrow composition. We can thus consider a second category $\mathcal{D}$ as being given by the objects $(F(A),F(B))$ and morphisms $F(f),F(g) \in \mathrm{Hom}_\mathcal{D}(F(A),F(B))$. The definition of functor just given encodes the important requirement that the map between categories it induces must be \textit{total}. That is, there must be no objects or arrows in $\mathcal{C}$ which are not mapped to at least one object or arrow in $\mathcal{D}$. 

Category theory has been applied in the context of theoretical structure because it is well-suited for the analysis of formal relations between theories. In particular, if one treats theories as collections of models together with a set of structure-preserving maps between the models (i.e. the morphisms are the isomorphisms of the theory) then it is natural to identify theories with categories and inter-theory relations (so longs as they are total) as functors. The idea is then that by characterising theories as categories we gain access to precisely the formal tools needed to enact comparisons between theories in terms of various `levels' of structure. Most relevant for comparison between physical theories are three cases that we will label \textit{equivalence}, \textit{surplus structure}, and \textit{surplus representational capacity}.\footnote{Here we are slightly modifying the terminology of \cite{bradley:2020,bradley:2025} who initially, following \cite{baez:2009}, talk about the third case in terms of `stuff' but then later provide an interpretation in terms of `representational freedom'.} These notions are always to be understood as relations between two categories relative to a functor from the first to the second. The three cases correspond to distinct property sets of the functor. The properties are as follows. Consider a functor $F : \mathcal{C} \to \mathcal{D}$:
\begin{itemize}
\item $F$ is \textit{faithful} if for any $f,g : A \to B$ in $\mathcal{C}$, whenever
$F(f) = F(g)$, it is also the case that $f = g$; that is, $F$ is injective from
$\mathrm{Hom}_{\mathcal{C}}(A,B)$ to $\mathrm{Hom}_{\mathcal{D}}(F(A),F(B))$.
\item $F$ is \textit{full} if for every morphism $g : F(A) \to F(B)$ in $\mathcal{D}$, there
is some morphism $f : A \to B$ in $\mathcal{C}$ such that $F(f) = g$; that is, $F$ is
surjective from $\mathrm{Hom}_{\mathcal{C}}(A,B)$ to $\mathrm{Hom}_{\mathcal{D}}(F(A),F(B))$.
\item $F$ is \textit{essentially surjective} if for every object $B$ in $\mathcal{D}$, there is some object $A$ in $\mathcal{C}$ such that $FA$ is isomorphic to $B$ in $\mathcal{D}$.
\end{itemize} 
If a functor is full and faithful then the morphisms are in one-to-one correspondence. If a functor is essentially surjective we can match objects (up to isomorphism). Thus, the existence of a functor that is full, faithful and essentially surjective is a natural condition for us to understand to categories as equivalent. 

Following \cite{bradley:2020}, in the context of physical theories characterised as categories, equivalence in this sense can be understood in terms of the relevant functor preserving \textit{both} structure and representational capacity. We can see this as follows: First, assume, as we indicated earlier, that the morphisms are isomorphisms. It is then natural to understand the amount of structure that a theory contains as indicated by the size of the isomorphism class. Since isomorphisms are structure preserving maps the fewer isomorphisms the more structure. By definition, if two theories are related by a functor that is full then there is a surjection between the is isomorphisms of the first theory and the isomorphisms of the second. This means that the second theory must have at least as much structure as the first. 

Second, consider the subclass of isomorphisms that map a model to itself. That is, the automorphisms of the objects. These can further be split into the trivial automorphisms given by the identity element and the non-trivial automorphisms given by the class of transformations identified as non-trivial symmetries of the models. A theory with non-trivial automorphisms provides us with a means to represent the same things using different models. Hence, given that we can match objects up to isomorphism, i.e. the functor is essentially surjective, the size of the non-trivial automorphism class corresponds to a theory's representational capacity. By definition, if  theories are related by a functor that is faithful we know that the non-trivial automorphisms of the first must map into non-trivial automorphisms of the second. Given that the two theories are related by the a functor that is also essentially surjective then we can understand the theories to have identical representational capacity. Thus a functor between two theories that is full, faithful, and essentially surjective will preserve both structure and representational capacity. 
 
We can then straightforwardly distinguish the two further cases by weakening the conditions on the functor in two directions. Consider a functor that is faithful and essentially surjective but not full. This functor will preserve representational capacity but forget structure. That is, there will be some isomorphisms of the second theory that are not mapped to by any isomorphisms of first. Since a theory with more isomorphisms has less structure we can then understand the functor as forgetting structure. Then consider a functor that is full and essentially surjective but not faithful. The particular case that is important for our purposes is when we find that trivial automorphisms of the second theory that are mapped to by non-trivial automorphisms of first. This means that whereas the first theory will have the capacity to represent the same situation via at least two non-trivially related models, the second will only have a single model to represent the counterpart situation. Thus, the functor forgets representational capacity. We will consider an example of each of these types of forgetful functors in the context of symplectic reduction later following the account of following the account of \cite{bradley:2025}. 
The three cases are summarised in the Table \ref{tab:functorial-relations}.

\bigskip
\begin{table}[ht]
\centering
\caption{Functorial relations between theories and their interpretive significance}
\label{tab:functorial-relations}
\begin{tabular}{>{\raggedright}p{3.5cm} c c c >{\raggedright\arraybackslash}p{5cm}}
\toprule
 & \textbf{Faithful} & \textbf{Full} &
 \begin{tabular}[c]{@{}c@{}}\textbf{Essentially}\\ \textbf{Surjective}\end{tabular}
 & \textbf{Interpretation} \\
\midrule
Theories $\mathcal{C}$ and $\mathcal{D}$ are \textbf{Equivalent} relative to functor $F : \mathcal{C} \to \mathcal{D}$. 
& $\checkmark$ & $\checkmark$ & $\checkmark$
& Isomorphisms and objects (up to isomorphism) are in one-to-one correspondence between $\mathcal{C}$ and $\mathcal{D}$.  \\
\midrule
Theory $\mathcal{C}$ has \textbf{Surplus Structure} compared with theory $\mathcal{D}$  relative to functor $F : \mathcal{C} \to \mathcal{D}$.
& $\checkmark$ & $\times$ & $\checkmark$
& There are some isomorphisms of $\mathcal{D}$ that are not mapped to by any isomorphisms of $\mathcal{C}$. \\
\midrule
Theory $\mathcal{C}$ has \textbf{Surplus Representational Capacity} compared with theory $\mathcal{D}$  relative to functor $F : \mathcal{C} \to \mathcal{D}$
& $\times$ & $\checkmark$ & $\checkmark$
& There are trivial automorphisms of the second theory that are mapped to by non-trivial automorphisms of the first. \\
\bottomrule
\end{tabular}
\end{table}
\bigskip

\subsection{Symmetry and Invariance}

Our next goal is to provide a more formal representation of nomic structure together with a language for describing how other theoretical structures vary with respect to this structure. We can do this by introducing the \textit{Nomic-AIR Formalism} formalism developed in detail in \cite[\S9]{gryb:2024}. We will then relate this formalism to the category theoretic  presentation of theoretical structure just given. 

In our earlier largely informal discussion we set out in qualitative terms how one can understand nomic structures as providing us with a \textit{partitioning map} and a \textit{projection map}. In more formal terms, these maps can be explicitly defined as follows:
\begin{definition}
\label{partitioning map}
\index{partitioning map}
\textbf{Partitioning Map}, $n$, is nomic structure that partitions the space, $K$, of KPMs into the proper subspace, $D\subset K$, of DPMs and the subspace, $\neg D$, of non-DPMs such that $K = D \cup \neg D$. In general, such structure can be represented as a non-injective map
	\begin{equation}
		n: K \to D\,.
	\end{equation}
\end{definition}
\noindent The assumption that the map is non-injective is to ensure that unless the laws impose no constraints on the kinematical possibilities, there will be at least one KPM that is not a DPM.

\begin{definition}
The \textbf{Projection Map}, $\pi_N$ can be defined in four steps:
\begin{itemize} 
\item [i.] First, consider the function $e: D \to \mathbbm R$ such that two elements $x,y \in D$ are \emph{dynamically equivalent}, $x \sim y$, iff $e(x) = e(y)$, where $\sim$ is the dynamical equivalence relation. 
\item [ii.] Second, use this equivalence relation to define the \emph{dynamical equivalence classes} $[x] = \{ y\, |\, y \sim x,\, \forall y \in D \}$. 
\item [iii.] Third, designate the space of equivalence classes $\tilde D$ of $D$ the space of \emph{Distinct Dynamically Possible Models (DDPMs)}. 
\item [iv.] Finally we can consider the projection from the space of DPMs to the space of DDPMs:
	\begin{equation}
	\pi_N: D \to \tilde D\,.
	\end{equation}
\end{itemize}
\end{definition}
\noindent The projection map will necessarily be a surjective function since for every element of $\tilde D$ we will have that $\pi_N(D)$ is well-defined. When it is non-trivial it will also be non-injective since it will categorize at least two models as dynamically equivalent.  An important special case is when the members of the equivalence class produced by $\sim$ are related by a group action. Here, $e$ gives $D$ a principal fibre bundle structure for the corresponding group, where $\pi_N$ is the bundle projection, $[x]$ are the bundle's fibres, and $\tilde D$ is the base space. This is precisely the case that we will consider in the context of symplectic reduction later. 

Now assume the \textit{non-constitutive} structures (i.e. nomic, spatiotemporal or other) of a theory to be represented by a map with the domain a space, $U$, and the codomain a space, $V$, so that we have $\mS: U \rightarrow V$ or equivalently $\mS(u)=v$ for $v\in V$ and $u \in U$. Let us then consider endomorphisms on the space of DPMs that is transformations $\phi: D \to D$ or equivalently $d'=\phi(d)$ for $d,d'\in D$. We can then consider the \textit{pullback of a structure by an endomorphism}, that is we can define the precomposition map as follows $\phi_d^*\mS$ by transferring the action of a particular endomorphisms $\phi_i$ to a particular structure $\mS_a$ by pre-applying it to the latter's domain: $
\phi_d^*\mS = \mS\circ\phi_d =  \mS(\phi(d))$. Consider for a theory a theory, $\mT$ the space of transformations, $\Psi$, given by the endomorphisms $\psi_i$ of $K$:
\begin{equation}
	\Psi = \{ \psi_i,\, \forall i\, |\, \psi_i: K \to K\} \,.
\end{equation}
Assume that $\mT$ is further equipped with nomic structure $N$ playing the role of a partitioning function $n: K \to D$. The space of transformations, $\Phi$, is the space of all endomorphisms $\phi_i$ of $D$:
\begin{equation}
	\Phi = \{ \phi_i,\, \forall i\, |\, \phi_i: D \to D\} \,.
\end{equation}

The space $\Phi$ is partitioned into three non-overlapping subspaces, \Abs{\mS}, \Rel{\mS}, and \Inc{\mS}, which we define as follows:

\begin{definition}
\label{defnA}
\index{absolute subspace}
\textbf{Absolute Subspace}: \Abs{\mS}: the subspace of all $\phi_i \in \Phi$ such that the pullback of the structure under $\phi_i$ is trivial for all tokens of the structure: $\phi_i^* \mS_a = \mS_a, \forall a$. Such structure $\mS$ transforms \emph{invariantly}, and is in this sense is \emph{absolute} under $\phi_i$.
\end{definition}  

\begin{definition}
\label{defnR}
\index{relative subspace}
\textbf{Relative Subspace}: \Rel{\mS}: the subspace of $\phi_i \in \Phi$ such that the pullback of the structure under $\phi_i$ is well-defined and non-trivial: $\phi_i^* \mS_{a} = \mS_{b}$, for some $a\neq b$. Such structure $\mS$ transforms \emph{covariantly}, and in this sense is \emph{relative} under $\phi_i$.
\end{definition} 
\begin{definition}
\label{defnI}
\index{incomplete subspace} 
\textbf{Incomplete Subspace}: \Inc{\mS}: the subspace of $\phi_i \in \Phi$ such that the pullback of the structure under $\phi_i$ does not have a closed action on the tokens of the structure: there exists at least one $\mS_a$ s.t. $\phi_i^* \mS_a \neq \mS_{b}, \forall b$. Since it is not closed under the action of $\phi_i$, such structure $\mS$ does not transform in an appropriately well-behaved way, and in this sense is \emph{incomplete} under $\phi_i$.    
\end{definition} 

We assume that the the Absolute-Incomplete-Relative (AIR) regions are non-overlapping (i.e. mutually exclusive), that the union of the three AIR regions must be $\Phi$ itself, and that the identity $\mathbbm 1$ is an element of \Abs{\mS}. It is then possible to prove that  each AIR region can be defined via the complement of the other two \cite[Theorem 9.1]{gryb:2024}. Furthermore, we can then introduce the following three further definitions:

\begin{definition}
\label{defn6}
\index{fixed structure!AIR definition}
\textbf{Fixed Structure}. The collection of all structures for some theory theory $\mT$, for which the relative subspace of $\Phi$ is empty:
\begin{displaymath}
 \Rel{\mS} = \emptyset.
\end{displaymath}
\end{definition}  
\begin{definition}
\label{symmetry}
\Abs{\pi_N}: \textbf{Symmetry transformations.} The subspace of all $\phi_i \in \Phi$ such that the pullback of $\pi_N$ under $\phi_i$ is trivial: $\phi_i^* \pi_N = \pi'_N=\pi_N$. Such transformations necessarily leave the space $\tilde D$ of DDPMs invariant. These are naturally interpreted as symmetry transformations of the theory.
\end{definition}
\begin{definition}
\label{dynamical}
\Rel{\pi_N}: \textbf{Dynamical transformations.} The subspace of all $\phi_i \in \Phi$ such that the pullback of $\pi_N$ under $\phi_i$ is well-defined and non-trivial: $\phi_i^* \pi_N = \pi'_N\neq\pi_N$. The transformations in this region induce diffeomorphisms of $\tilde D$ that transform between DDPMs. These are naturally interpreted as the dynamical transformations of the theory.
\end{definition}
\noindent An important special case is when the equivalence classes of $D$ are identical manifolds $\mathcal F$ so that the projection $\pi_N$ equips $D$ with a fibre bundle structure. In this case the symmetry transformations $\phi_v \in \Abs{\pi_N}$ are the vertical directions in the bundle and the dynamical transformations $\phi_h \in \Rel{\pi_N}$ are the horizontal directions. As already noted, fixed in our sense is a weaker requirement than that in the sense of \cite{pooley:2017} since it places no requirement on invariance of structures across KPMs. Rather, fixed in our sense is essentially identical to idea of a `absolute field' introduced in \cite[Def. 8]{read:2024}. 

These ingredients are sufficient to provide a precisification of the standard notions of invariant structure and surplus structure via respective behaviour under symmetry transformations. An \textbf{Invariant Structure} is absolute (and therefore invariant) under all symmetries of $\mT:\Abs{\mS} \supseteq \Abs{\pi_N}$ and \textbf{Surplus Structure}: $\mS$ transforms in a well-behaved manner under \textit{some} of the symmetries of $\mT: \Rel{\mS} \cap \Abs{\pi_N} \neq \emptyset$. Correspondingly, we can also  provide a precisification of the notions of a structure that is absolute or relative via respective behaviour under dynamical transformations. A \textbf{Dynamically Absolute Structure} does not vary under some transformations between DDPMs and when $\mS$ has non-trivial relative transformations, these are always symmetry transformations. Such structure \textit{never} transforms as one varies between DDPMs (see \cite[Def. 9.18]{gryb:2024} for formalisation). A \textbf{Dynamically Relative Structure}: $\mS$ transforms as one varies between some DDPMs: $\Rel{\mS} \cap \Rel{\pi_N} \neq \emptyset$. It is then straightforward to prove that invariant structures and surplus structures are mutually exclusive \cite[Theorem. 9.3]{gryb:2024}, dynamically absolute and dynamically relative structure are mutually exclusive \cite[Theorem. 9.4]{gryb:2024}, invariant structure that is dynamically absolute is always fixed structure \cite[Theorem. 9.7]{gryb:2024} and when a structure is both surplus and dynamically absolute then all relative transformations of the structure will be symmetry transformations \cite[Theorem. 9.8]{gryb:2024}.

\subsection{State Spaces as Objects}

We can now turn the project of relating this formalism to that in terms of category theory provided earlier. The most basic ambiguity in relating the two presentations is the identification one draws between the `models' that are the objects within the categories and the sets of kinematically and dynamically possible models that are picked out by the constitutive and nomic structures. Practice is not entirely consistent since in the context of spacetime theories it is standard to take both the models qua category theory objects and the models qua DPMs to correspond to spacetimes. For example, general relativity might be understood as the category defined by specifying Lorentzian manifolds (together with a suitable set of functions on them) as objects and isometries as morphisms. The nomic structure of the theory is then given by the Einstein field equations which pick out pairings of metric and stress energy tensors as the DPMs which \textit{are} the Lorentzian manifolds in question. 

By contrast, in the context of theories in initial value formulation, it is standard to take the models qua category theory objects as entire state spaces with the models qua DPMs as state space curves. For example, Hamiltonian mechanics might be understood as the category defined by specifying symplectic manifolds (together with a suitable set of functions on them) as objects and symplectomorphisms as morphisms. By contrast, the nomic structure is given by Hamilton's equations which, for a regular theory, pick out a unique Hamilton vector field the integral curves of which are the DPMs. This dual use of the term model creates no inherent problem so long as we keep track of what we are talking about when we talk about the `models' or `structure' of a theory. In particular, it is natural to understand the objects in a category theoretic  formulation of an initial value theory as \textit{state spaces} containing the DPMs for specific tokens of constitutive and nomic structure. Thus in specifying tokens of the constitutive and nomic structure we are specifying the objects of the salient category. 

The immediate implication is then that we can interpret the symmetry transformations as the morphisms of the category. That is, since these are precisely the maps that preserve the token of the nomic structure they will leave both the state spaces and $\tilde D$ invariant. For the case at hand they will be invertible and thus isomorphisms. Let us indicate this identification via a compressed notation where the objects are indicated by triples $\langle K,\pi_n,\mathcal{S}\rangle$, understood given by tokens of the constitutive and nomic structure together with some further set of structures (e.g. matter or geometric structure) and the isomorphisms by $\Abs{\pi_N}$ as per our definitions above. We can then consider functors between categories as corresponding to maps between: i) the same tokens of constitutive and nomic structure but different tokens of some further set of structures, e.g. solutions of a 3-body Newtonian model with different initial positions; ii) different tokens of  constitutive, nomic structure, further structure, e.g. 2-body and 3-body Newtonian models, each for specific initial positions; or iii) tokens of \textit{different} constitutive and nomic structure and further structure e.g. classical and quantum mechanics Harmonic oscillators. 

The intersection between these two presentations is a rich terrain that provides a powerful toolkit to disambiguate various philosophical and mathematical issues in the interpretation of physical theory. A straightforward and fruitful task is to translate the  notions of equivalence, surplus structure, and surplus representational capacity defined above into the Nomic-AIR terminology. We will do this by considering their meaning in the context of a functor that maps between the categories as we have just defined them. In particular, consider a first category $\mathcal{C}$ as given by objects $A^i=\langle K^1,\pi^1_n,\mathcal{S}^1\rangle$, $B=\langle K^2,\pi^1_N,\mathcal{S}^1\rangle$ and isomorphisms $\Abs{\pi^i_N}$, where $i$ runs over all the tokens or types in question. We can then define a functor $F: \mathcal{C}\rightarrow\mathcal{D}$ to a second category $\mathcal{D}$ with objects $F(A)=\langle F(K^1),F(\pi^i_N),F(\mathcal{S}^1)\rangle$, $F(B)=\langle F(K^2),F(\pi^2_N),F(\mathcal{S}^2)\rangle$  and isomorphisms $\Abs{F(\pi^i_N}))$. 

It is then straightforward to interpret the three functional relations. Essential surjectivity holds in all three cases and indicates that we can match objects (up to isomorphism) between the two categories. In our case the objects are families of DPMs for specific tokens of constitutive and nomic structure together with further additional structures such as matter structure. Equivalence then means that the functor is also full and faithful which means that we can match the isomorphisms as well. In our case the isomorphisms are symmetry transformations. Thus equivalence between two theories indicates that for every family of DPMs together with associated constitutive and nomic structure together with further additional structures and symmetries in $\mathcal{C}$, there is a corresponding family of DPMs together with associated constitutive and nomic structure together with further additional structures and symmetries in $\mathcal{D}$. 

Surplus structure is the case in which the functor is not full and thus there are some some isomorphisms of $\mathcal{D}$ that are not mapped to by any isomorphisms of $\mathcal{C}$. This means that there are  symmetries of $\mathcal{D}$ that are not symmetries of $\mathcal{C}$ and thus the space of DDPMs in $\mathcal{D}$ will be strictly smaller than the space of DDPMs in $\mathcal{C}$. The most important example of such a functor is given by specifying the structures of some theory with some symmetry class and then constructing a new theory based on the first by enlarging the symmetry class. Such enlargement may be merely by stipulation under some symmetry principle, such as all DPMs related by variational symmetries are DDPMs, without changing the nomic structure of a theory. One might also consider cases where we explicitly change the nomic structure to build in more symmetry, such as for instance in moving from Newtonian mechanics to Barbour-Bertotti theory \citep{barbour:1982}. In canonical terms such a move corresponds precisely to enlarging the symmetry class by moving from a regular to an irregular representation of nomic structure,  see \cite[p.168]{gryb:2024}.

Finally, we can consider the case of surplus representational capacity. This corresponds to the functor being full and essentially surjective but not faithful. The most important case  for our purposes is when there are trivial automorphisms of $\mathcal{D}$ that are mapped to by non-trivial automorphisms $\mathcal{C}$. Such a functor is given by considering a theory and projecting out symmetries by moving to a reduced space of DPMs. This is removing representational capacity by redefining the space of DPMs such that distinct symmetry related DPMs in $\mathcal{C}$ get mapped to the same DPM in $\mathcal{D}$. The most extreme example of such a symmetry reduction would be when the absolute subspace of the projection map of the nomic structure consists solely of the identity element \cite[p.155]{gryb:2024}. Such a `categorical' theory would have \textit{no} surplus representational capacity since there would be only one dynamically distinct model . These relationships are summarised Table \ref{tab:functorial-relations2}.

\begin{table}[ht]
\centering
\caption{Functorial relations between theories in Nomic-AIR formalism.}
\label{tab:functorial-relations2}
\begin{tabular}{>{\raggedright}p{3.5cm} c c c >{\raggedright\arraybackslash}p{5cm}}
\toprule
 & \textbf{Faithful} & \textbf{Full} &
 \begin{tabular}[c]{@{}c@{}}\textbf{Essentially}\\ \textbf{Surjective}\end{tabular}
 & \textbf{Interpretation} \\
\midrule
Theories $\mathcal{C}$ and $\mathcal{D}$ are \textbf{Equivalent} relative to functor $F : \mathcal{C} \to \mathcal{D}$. 
& $\checkmark$ & $\checkmark$ & $\checkmark$
& For every family of DPMs together with associated constitutive and nomic structure together with further additional structures and symmetries in $\mathcal{C}$, there is a corresponding family of DPMs together with associated constitutive and nomic structure together with further additional structures and symmetries in $\mathcal{D}$.  \\
\midrule
Theory $\mathcal{C}$ has \textbf{Surplus Structure} compared with theory $\mathcal{D}$  relative to functor $F : \mathcal{C} \to \mathcal{D}$.
& $\checkmark$ & $\times$ & $\checkmark$
& There are symmetries of $\mathcal{D}$ that are not symmetries of $\mathcal{C}$ and thus the space of DDPMs in $\mathcal{D}$ will be strictly smaller than the space of DDPMs in $\mathcal{C}$\\
\midrule
Theory $\mathcal{C}$ has \textbf{Surplus Representational Capacity} compared with theory $\mathcal{D}$  relative to functor $F : \mathcal{C} \to \mathcal{D}$
& $\times$ & $\checkmark$ & $\checkmark$
& There are non-trivial automorphisms in $\mathcal{C}$ which get mapped to the identity in $\mathcal{D}$ and thus there are distinct symmetry related DPMs in $\mathcal{C}$ which get mapped to the same DPM in $\mathcal{D}$. \\
\bottomrule
\end{tabular}
\end{table}

\section{Irregularity and Ill-posedness}
\label{Irregularity}

The distinction between the cases of regular and irregular nomic structure is important in the context of both classical and quantum theories. The relevant criterion is more straightforward to identify in the classical context where it can be directly connected to the well-posedness of the evolution equations. In the context of Lagrangian action principles this is a distinction which can be drawn between \textit{regular Lagrangians} and the class of \textit{irregular Lagrangians with initial value constraints}. This distinction in turn maps onto that between unconstrained and constrained Hamiltonian theories with the important exception of totally constrained systems, which in general have \textit{irregular Lagrangians without initial value constraints} cf. \cite[\S 11]{gryb:2024}. In this section we will provide an interpretative overview of irregular nomic structure focusing on the connection between irregularity and Hamiltonian constraints, Dirac's Theorem which connects Hamiltonian constraints to redundancy, and the geometric conditions for an irregular Lagrangian to be associated with initial value constraints via distinction between transverse and tangential orientation of the irregular variations. We will make reference to some the ideas  from the previous section where relevant as the next step in our synthetic project. For further details and the explicit examples see  \cite[\S7-8]{gryb:2024}.\footnote{See also \cite{sudarshan:1974,sundermeyer:1982,Henneaux:1992a,diaz:2014,diaz:2018,diez:2020}. See also \cite{Dirac:1958a,Dirac:1964}. See \cite{Salisbury:2017} for historical analysis highlighting the role of L\'{e}on Rosenfeld. }

\subsection{Regular and Irregular Nomic Structure}
\label{RINS}

Let first review some standard but important formal material that  establishes the connection between irregular Lagrangians and  Hamiltonian constraints. Consider a finite dimensional system with action:
\begin{equation}
I=\int_{t_1}^{t_2}\de t\, L(q^i,\dot{q}^i,t)
\end{equation}
where the index $i=1,...,N$ runs over the degrees of freedom of the system and write the Euler--Lagrange equations in the form:\index{Euler--Lagrange operator}
  \begin{equation}
  \label{Euler}
 \alpha_i (q,\dot{q},t) := \frac{\partial L}{\partial q^i} -  \frac{\partial}{\partial t}\frac{\partial L}{\partial \dot{q}^i} = W_{ij}(q, \dot{q})\ddot{q}^j - K_i =0
 \end{equation}
  where  $\alpha_i$ is the Euler--Lagrange operator, $W_{ij}$ is the Hessian matrix:\index{Hessian matrix}
\begin{equation}
\label{Hessian}
 W_{ij}  = \frac{\partial^2 L}{\partial \dot{q}^i \partial \dot{q}^j}  
\end{equation}
and 
\begin{equation}
\label{kequation}
K_i=\frac{\partial L}{\partial q^i}-\frac{\partial^2 L}{\partial\dot{q}^i \partial q^j} 
 \dot{q}^j\,.
\end{equation}
Writing the Euler--Lagrange equations in this form makes clear that the accelerations \textit{at any particular time} are uniquely determined by the velocities \textit{at that time} if and only if the Hessian can be inverted. This is equivalent to the condition that the determinate of the absolute value of the Hessian is non-zero. 
We define a \emph{regular} Lagrangian as a Lagrangian that leads to Euler--Lagrange equations that can be uniquely determined in terms of the initial configurations and velocities, and therefore satisfies $\text{det}\lvert W_{ij} \rvert\neq 0$. This implies that the Euler--Lagrange equations are autonomous, integrable and of second-order. Furthermore, if the Lagrangian is regular then we are guaranteed by the inverse function theorem that there exists a one-to-one transformation between the configuration-velocity variables $(q,\dot{q},t)$ and the configuration-momentum variables $(q,p,t)$. This, in turn, means that the coordinates and momenta are independent variables \cite[p. 11]{sundermeyer:1982}. Systems with regular Lagrangians have regular nomic structure in the sense that the equations of motion, both the second-order Euler--Lagrange equations and the first-order Hamiltonian equations, will be well-posed for all degrees of freedom. 

Theories with regular nomic structure are thus precisely those unconstrained Hamiltonian theories which the transformation to the phase space variables is guaranteed to be unique and the coordinates and momenta will then be independent variables. Following our discussion of theoretical structure in the previous section, for unconstrained Hamiltonian theories it is natural to understand the models as being picked out by a triple $\langle\Gamma,\mathcal{A},\omega,\rangle$ consisting of the even-dimensional symplectic manifold $\Gamma$, the Poisson algebra  $\mathcal{A}$ of smooth, real functions on $\Gamma$, and the symplectic two-form that equips both the manifold and algebra with the salient geometric and algebraic structures. Models defined via these structures are sufficient to determine (regular) nomic structure via Hamilton's equations since they allow us to define for any observable quantity and Hamiltonian function, $A, H\in\mathcal{A}$, time evolution automorphism via the Poisson bracket, $\dot{A}=\{A,H\}:=\omega(X_A,X_H)$ where $X_A$ and $X_H$ are the Hamilton vector fields induced by the symplectic structure on phase space via $\iota_{X_A}\omega = dA$ which is guaranteed to always have unique solutions since $\omega$ is non-degenerate. This is the hallmark of a Hamiltonian theory with regular nomic structure: Hamilton's equations encode the nomic structure and pick out a unique set of internal curves on phase space as the DPMs. The isomorphisms are then given by diffemophisms whose push forward preserves both the symplectic and Poisson algebra structure. These are the \textit{symplectomorphism} $\phi$ which are by definition such that $\phi^* \omega' =\omega$ and will also be guaranteed to be such that $\phi^*\mathcal{A} \rightarrow \mathcal{A}$. 

Let us then define an \textit{irregular} (or \textit{singular})  Lagrangian as a Lagrangian for which $\text{det}\lvert W_{ij} \rvert = 0$. This implies that at least some of the components, $\ddot q^i$, of acceleration at a given time cannot be solved uniquely in terms of the configuration and velocity variables at that time \cite[p. 39]{sundermeyer:1982}. We can represent the breakdown of integrability via the null space of the Hessian.  Furthermore we can show that each dimension in the null space of the Hessian will correspond to an independent relations between canonical variables a Hamiltonian constraint. Explicitly, for an $N$ dimensional system, if we have that the rank of the Hessian is $R < N$, then we can consider the $N-R$ null eigenvectors $\lambda_{(a)}^i  (q, \dot{q})$:
\begin{equation}
\label{Hessker}
\lambda_{(a)}^i (q, \dot{q}) W_{ij}(q, \dot{q}) =0
\end{equation}
for $i,j=1,...,N$ and $a=1,...,N-R$. This should be compared to the case of a regular Lagrangian system where the non-null directions are used to solve the Euler--Lagrange equations by integrating \eqref{Euler}. We can introduce a set of explicit \textit{canonical constraints} which describe the relations between positions and momenta via a relation of the form:
\begin{equation}
\label{explicitconstraints}
\varphi_s=p_s-g_s(q,p_\alpha)=0
\end{equation}
where the index $s=1,...,N-R$ runs over the canonical variables which are not independent,  $\alpha=1,...,R$ runs over the remaining independent variables, and the functions $\varphi_s$ encode the relevant interdependencies. These relations are called \textit{primary constraints} following the coinage of \cite{anderson:1951}. From these relations we can, at least locally, derive an explicit form of the dynamics of a constrained Hamiltonian by substituting  (\ref{explicitconstraints}) to eliminate the $N-R$ dependent momenta, deriving the analogue of Hamilton's equations, and then applying consistency conditions leading to \textit{secondary constraints}. Applying consistency conditions again may lead to further constraints and it is only given a termination of this procedure that a well-posed first order formalism can be established. For the most basic case the primary constraints are automatically propagated consistently and all constraints are primary constraints. In what follows we will also assume that all constraints are \textit{first-class}, which means that they are mutually Poisson commuting on the sub-manifold that they define on phase space, that the constraint algebra is a Lie algebra, which means that their Poisson bracket relations close with at most structure constants, and that they Poisson commute with the Hamiltonian.\footnote{For more details on the explicit approach  see \cite[pp. 45--50]{sundermeyer:1982} and \cite[pp, 13--16]{gitman:1990} and \cite{Henneaux:1992a} for the implicit approach, which is more general, see \cite{sundermeyer:1982}.} 

There are two natural methods for the category theoretic  presentation of the models and isomorphisms of a constrained Hamiltonian theory: extrinsic and intrinsic. In the extrinsic, we consider the theory as formulated on the full phase space in the intrinsic we define the salient objects directly on the constraint manifold. Following the proposal of \cite{bradley:2025} (made in a slightly different context) we can formulate the extrinsic category \textbf{ConHam} as having its objects the models as picked out by the quadruple  $\langle\Gamma,\mathcal{A},\omega,\varphi_s\rangle$, where the $\varphi_s$ are the full collection of constraints which are (by assumption) first class and primary in the terminology introduce above, and the isomorphisms as given by the subclass of symplectomorphism that also preserve the constraints.\footnote{\textbf{UnconHam} is the analogous to the category \textbf{TotHam} in \cite{bradley:2025}  in that it also defines a Hamiltonian theory not restricted to the final constraint surface. However, formally the two categories are quite different in that in \textbf{TotHam} the relevant models are defined on the (first class) primary constraint surface of theory with secondary constraints.} That is, the diffemophisms $f: \Gamma \rightarrow \Gamma$ which are such that their push-forward satisfies $f^* \omega' =\omega$, $f^*\mathcal{A} \rightarrow \mathcal{A}$, and $f^*\varphi_s \rightarrow \varphi_s$. We might then understand the dynamics to be defined simply by Hamilton's equations as before. So we could again define expressions of the form: $A, H\in\mathcal{A}$, $\dot{A}=\{A,H\}:=\omega(X_A,X_H)$, and $\iota_{X_H}\omega = dH$. \textit{However, this nomic structure is pathological since it does fulfil the partition function}. Although the Hamilton vector fields $X_H$ and $X_A$ are unique they do not lie everywhere along the constraint surface and thus supposedly dynamical trajectories in phase space may start off by obeying the constraints and then move away from the surface. 

The obvious response to this pathology is to move to an intrinsic presentation where we seek to construct models explicitly within the sub-manifold $\Sigma$ defined by the constraints. Let us define the  inclusion map  $i : \Sigma \hookrightarrow \Gamma$ and then push this map forward onto the symplectic structure to define a two form $\tilde{\omega} =  i^* \omega$ and onto the algebra of phase space functions to define a sub-algebra of functions that are restricted to the constraint surface, $\tilde{\mathcal{A}}= i^* \mathcal{A}$ where the binary operation is given by $\tilde{\omega}$. Since our consistency conditions imply that
$\{\varphi_i,\tilde{H}\}_\Sigma := \tilde{\omega}(X_{\varphi_i},X_{\tilde{H}})=0$,
we are guaranteed that the dynamics defined by
$\iota_{X_{\tilde{H}}}\tilde{\omega} = d\tilde{H}$
and
$\dot{A}_{\Sigma}:= \omega(X_{\tilde{A}},X_{\tilde{H}})$  will be restricted to the constrain surface. However, crucially, the two form $\tilde{\omega}$ will in general be closed but degenerate. It is a pre-symplectic two form and has a non-trivial kernel consisting of the null vector fields $\iota_{X_{\varphi_i}}=0$ associated with the constraints. This means that the dynamical equations are not uniquely determined. \textit{This nomic structure is still pathological since it does fulfil the projection function}. We will return to this issue shortly. 

Following the proposal of \cite{bradley:2025} we can then formulate the intrinsic category \textbf{PreHam} as having its objects the models as picked out by the tuble $\langle\Sigma,\tilde{\omega},\tilde{\mathcal{A}}\rangle$ and consider the isomorphisms defined via the diffemophisms $g: \Sigma \rightarrow \Sigma $ which are such that $g^*\tilde{\omega}'=\tilde{\omega}$ and $g^*\tilde{\mathcal{A}}'=\tilde{\mathcal{A}}$.\footnote{This is the category that \cite{bradley:2025} calls \textbf{ExtHam} with only notational changes.}  One can insightfully apply the tools for the analysis of theoretical structure via conditions on functors to our two presentations of the constrained Hamiltonian formalism. Once more following \cite{bradley:2025}, we can define a functor $F:\textbf{ConHam} \rightarrow \textbf{PreHam}$ that takes each model $\langle\Gamma,\mathcal{A},\omega,\varphi_s\rangle$ to its restriction to the points that satisfy the constraints $\varphi_s=0$, i.e. the associated model $\langle\Sigma,\tilde{\omega},\tilde{\mathcal{A}}\rangle$, and takes each isomorphism $f$ to its action on $\Sigma$. Proposition 2 of \cite{bradley:2025} then takes the form:

\begin{proposition}
 $F:\textbf{ConHam} \rightarrow \textbf{PreHam}$ is faithful and essentially surjective but not full. 
\end{proposition}

\noindent The proof is identical to that provided in \cite[A.1-2]{bradley:2025} mutatis mutandis.\footnote{The difference between the two cases is entirely with regard to the the fact that our \textbf{ConHam} is defined via symplectic structure whereas Bradley's \textbf{ExtHam} is defined via presymplectic structure. The structure of the proof is then such that it will carry across between the two cases.} Following the interpretational approach summarised in Table \ref{tab:functorial-relations}, we can understand the move to the constraint surface as the elimination of surplus structure. There are isomorphisms  of  \textbf{ConHam} that are not isomorphisms of \textbf{PreHam} in particular transformations that correspond to moving along the null vector fields $X_{\varphi_i}$. 

Significantly, however, we \textit{cannot} at this point move to an interpretation of these theories within the Nomic-Air formalism according to Table \ref{tab:functorial-relations2} since we do not have a sufficient formal basis to identify the DPMs in  \textbf{ConHam} (due to the lack of partition) or the DDPMs in \textbf{PreHam} (due to the lack of projection). The problem is that in both cases the nomic structure is pathological: it does not provide us with a well-posed initial value (or boundary) problem. For \textbf{ConHam} the problem is obviously related to the failure to restrict to the constraint surface and that is precisely what moving to \textbf{PreHam} solves. However, whilst understanding a constrained Hamiltonian theory via this category might appear to provide us with the right level of structure to pick out solutions up to isomorphism, since the salient nomic structure is itself only defined up to isomorphism it is formally ill-posed as an initial value problem precisely because it fails to project out a \textit{pernicious} form of redundancy between DPMs, cf. \cite[p.14]{bradley:2025}.\footnote{This issue has been discussed a number of time in the philosophical literature on constrained Hamiltonian without its general relevance to debates on symmetry seemingly being full appreciated. See \cite{Belot:2003,Thebault:2011b}. } The following two-subsections are devoted to the diagnosis of this issue following, first, the traditional Dirac approach and, second, the geometric approach of \cite{gryb:2024}. We will return to the problem of identifying a non-pathological presentation of the nomic structure in Section \ref{Bradley} in the context of the discussion of \cite{bradley:2025} on symplectic reduction. 


\subsection{Dirac's Theorem}\label{sec:Dirac thm}

The first step is to consider explicitly the connection between the null-vector fields associated with the constraints and the ill-poedness of the initial value problem. Following a highly influential argument due to \cite{Dirac:1964}, it is possible to prove, in certain restricted circumstances, a direct correspondence between the directions associated with the first class of constraints within a Hamiltonian formalism. If we restrict to theories in which the constraints automatically propagate consistently (they are `primary constraints') and have vanishing Poisson bracket with each other (they are `first class constraints') then it is possible to prove what is often called `Dirac's Theorem' and  which can be presented more formally as follows:

\begin{proposition}
Given that:
\begin{itemize}
\item [i.] We have a constrained Hamiltonian theory with first-class primary constraints $\phi_a$, with associated \textit{totally arbitrary} multipliers $v^a$, a time dependent phase space function $F(t): (q,p)_t \rightarrow \mathbb{R}$, and \textit{external time parameter}, $t$.
\item [ii.] The physical state of the system at an initial time $S_{t}$ can be specified  by a full set of canonical variables $(q,p)_{t_1}$; that is, although we must allow for physical states to \textit{not} be uniquely determined by specified of canonical variables, we can assume that canonical variables uniquely determine physical states.  $S_{t}$ thus \textit{supervenes} on $(q,p)_{t}$.
\item [iii.] Whenever we make a specification of the physical state at an initial time, the equations of motion then fully determine the physical state at other times. 
\item [iv.] Define any transformation of the canonical variables that does not change the physical state as a \textit{gauge transformation}. 
\end{itemize}
Then, 
\begin{itemize}
\item [D.] The first-class primary constraints, $\phi_a$, generate gauge transformations.
\end{itemize}
\end{proposition}
\noindent See \cite[p.117]{gryb:2024} for the proof after the original argument of \cite{Dirac:1964}. The implications of Dirac's theorem are easy to understand: we should treat as dynamically distinct only those solutions that are independent of the flows on phase space associated to the first-class primary constraints and, conversely, phase space points that lie along the orbit of a first-class primary constraint should be taken to represent physically identical states of affairs. This is precisely an interpretive justification of the geometric definition the isomorphism class in \textbf{ConHan} above. 
 
There are two important restrictions to the scope of Dirac's theorem which limit its applicability. First, in general, whilst we \textit{can} expect most physical theories to feature constraints that are first-class and not second class, many theories, such as electromagnetism and general relativity, feature constraints that are secondary as well as primary. In his original presentation Dirac makes the `conjecture' that secondary first-class constraints might also generate transformations of the physical variables that do not change physical states. There is then a large literature discussing the extent to which the conjecture can in fact be proved, and the relevance of `pathological' counter-examples.\footnote{See \cite{lusanna:1990a,lusanna:1991,Pons:2005} and the more complete lists of references therein. Recent philosophical discussion can be found in \cite{Pitts2014b,Pitts2022,Pitts2024,PooleyWallace2022,Bradley2025a,Bradley2025b,Bradley2025c}.} Second, as notably discussed by \cite{Barbour:2008}, the restriction to theories with an external time parameter is very limiting. In particular, any theory which is time reparametrization invariant will fall outside the scope of Dirac's theorem. In such `totally constrained' theories the Hamiltonian is itself a constraint and the association between the constraint action and gauge transformation leads to the infamous problem of time.\footnote{See \cite{Kuchar:1991,isham:1993,pons:2010,anderson:2017,Gryb:2016a,gryb:2024,CASADIO2024169783}} 

Over and above these formal restrictions the conceptual weakness of Dirac's approach is that it involves introducing a notion of gauge transformation both too general and too specific. It is too general since it opens up a vagueness as to what we mean by `does not change the physical state'. In particular, such a notion of gauge would, under some interpretations of the physical state, include rigid global transformations, like uniform spatial translations, which have nothing to do with first-class constraints nor ill-posed initial value problems.  It is too specific since it restricts us to considering dynamical redundancy for instantaneous states defined relative to an external temporal background. What we would really like is a notion of gauge transformation that is \textit{directly} connected to the irregularity of the Lagrangian and the fact that the accelerations depend on arbitrary functions and their derivatives \cite[p. 89] {sundermeyer:1982} and moreover makes explicit precisely when the existence of constraints derives from ill-defined initial value problems. This calls for a more geometric approach to the problem.

\subsection{Isochronous symmetries and Initial Value Constraints}\label{sec:VPS construction}

The first step in constructing a geometric formalism for the analysis of gauge symmetries in irregular Lagrangian systems is to consider a first-order Lagrangian form of the variational principle and equations of motion. This allows us to express the variational principle and symmetries in terms of geometric quantities on the tangent bundle. To adopt such an approach is follow the example of Emmy Noether rather than Paul Dirac. That is, unlike in the Dirac approach considered above, we assume an action, $S$, has a particular symmetry, then investigate its consequences. As such, the structure of the approach we will follow is in the same spirit as Noether's approach in deriving her two theorems, whilst differing on some crucial details and scope.\footnote{See \cite[p. 119-121, 125-128, 181]{gryb:2024} for detailed discussion of the role and limitations of Noether's theorem in the context of the analysis below. See \cite{lusanna:1991,kosmann:2010} and \cite[\S4.4]{Olver:2000} for more details on Noether's theorem and its relation to modern geometric analysis of symmetries and Lie groups.}

Building on the analysis of \cite[\S8]{gryb:2024} and \cite{gryb:2025}, which is partially based on \cite{woodhouse:1997}, we can introduce the basic elements of our formalism as follows. First, consider the action $S[\gamma] = \int_\gamma L(q^i, \dot q^i)$, which is first order in the configuration variables $q^i \in \mathcal C$, as a functional of the curve $\gamma: \mathbbm R \to \mathcal C$ parametrized by the time variable $t$.\footnote{Note that the generalisation to higher order actions is straightforward and that $i$ can be considered a continuous index for field theory applications.} Then, define the \textit{velocity phase space} $\mathcal{V} = T\mathcal C$ equipped with local coordinates $(q^i, v^i)$ and treat the Lagrangian $L(q^i, v^i)$ as a function on this space. We can write a first-order action, $S_1[\gamma]$, which is equivalent to but distinct from the second-order action $S$, using the one-form
\begin{equation}\label{eq:theta}
	\theta = \diby{L}{v^i} \de q^i\,
\end{equation}
and the Hamiltonian function 
\begin{equation}\label{eq:hamiltonian}
	H = v^i \diby L {q^i} - L\,.
\end{equation}
Note that, unlike on phase space proper, $\theta$ and $H$ are explicit functions of the Lagrangian $L$. Using these quantities, we have that
\begin{equation}
	S_1[\gamma] = \int_{t_1}^{t_2} \de t \lf( \iota_X \theta - H \rt) = \int_\gamma \lf( \theta - H \de t \rt)\,.
\end{equation}
The vector field
\begin{equation}\label{eq:X def}
	X := \dot \gamma = \dot q^i \diby{}{q^i} + \dot v^i \diby{}{v^i}\,
\end{equation}
is the tangent to a trial curve $\gamma$, now a path in $\mathcal{V}$, at time $t$, where dots are time derivates with respect to $t$ and fields are taken to over the extended velocity phase space $\mathcal{V}_t = \mathbbm R \times \Gamma$.

Let us now consider variations of the action $S_1$ with respect to a vector field $u \in \mathbbm R \times T\mathcal{V}$ satisfying $\iota_u \de t = 0$, where $\de t$ is understood as the differential induced by the embedding of $\gamma$ into $\mathcal{V}$. Since $\iota_X \de t = 1$, this condition implies that $u$ has no component along the tangent $X$ to $\gamma$, and therefore can't reparameterise $\gamma$. We will call such variations \emph{isochronous} since they preserve the time parameterization. Note that this does \emph{not} mean that $u$ must be time independent: as a function of $\mathbbm R$ it can vary in time while having no component along $X$ as a vector in velocity phase space. The important case where arbitrary parameterizations of $\gamma$ are allowed will be treated separately in the context of reparameterization invariant mechanics and the problem of time, see  \cite[\S12]{gryb:2024} and \cite{Gryb:2026c}.

Using this assumption, a straightforward (and well-known) calculation shows that the variational principle $\Lie_u S_1 = 0$, with $u(t_1) = u(t_2)=0$, leads to Hamilton's equations
\begin{equation}\label{eq:Ham eqns}
	\iota_X \omega + \de H = 0\,
\end{equation}
in their usual geometric form but now on velocity phase space so that $\omega = \de \theta$ and $H$ are taken to be functions on $\Gamma$. Equivalence to the second-order formalism can be seen by noting that the $\de v^i$ leg of this equation reduces to Hamilton's first equation $\dot q^i = v^i$ while the $\de q^i$ leg then reduces to the Euler-Lagrange equations arising from $S$.

Let us now study the consequences resulting from $S_1$ containing symmetries. Instead of considering, as above, arbitrary isochronous variations generated by $u$ and looking for conditions on $X$ that satisfy $\Lie_u S_1 = 0$, we consider \emph{particular} isochronous $u_\alpha$ and take if for granted that $\Lie_{u_\alpha} S_1 = 0$ for \emph{arbitrary} $X$ (i.e., even $X$ not satisfying Hamilton's equations). We take this to be the condition to be satisfied for the action $S_1$ to have symmetries.

Let us start by assuming that the symmetries of the second-order Lagrangian theory take the form
\begin{equation}\label{eq:q var}
	\delta_\epsilon q^i = T^{i}_{\alpha} \epsilon(t)\,,
\end{equation}
where $T^i_\alpha$ are operators on configuration space only (i.e., they are independent of the velocities $\dot q^i$) generating the symmetries and $\epsilon(t)$ is a time-dependent gauge parameter. For a gauge group $G$ with structure constants $f^\gamma_{\alpha\beta}$, the generators are representation of the algebra $\mathfrak g$ of $G$:
\begin{equation}\label{eq:T vec alg}
	[\bT_\alpha, \bT_\beta] = f^\gamma_{\alpha\beta} \bT_\gamma\,,
\end{equation}
in terms of the vector fields $\bT_\alpha = T^i_\alpha \partial_i$ in $T\mathcal C$. This is a restricted form of gauge transformations as compared to the normal Noether setup. As we will see, this restriction limits our attention to theories with only primary constraints, consistent with the assumptions of this paper. The more general case is treated in \cite{Gryb:2026}, cf. \cite[\S8.4]{gryb:2024}.

The symmetries \eqref{eq:q var} act on velocity phase space through the \emph{prolongation} vector field
\begin{equation}\label{eq:prolongation}
	u_\alpha = \epsilon \lf( T^i_\alpha \partial_{q^i} + \dot T^i_\alpha \partial_{v^i} \rt)  + \dot \epsilon \lf(T^i_\alpha \partial_{v^i}\rt) \,,
\end{equation}
which can be found by differentiating \eqref{eq:q var} and reading off the terms of the $q^i$ and $v^i$ variations. Note that $u_\alpha$ splits into two terms multiplying different time derivatives of $\epsilon$. These terms are independent at a given $t$. Let us call them
\begin{align}\label{eq:ui}
	u_{0,\alpha} &= T^i_\alpha \partial_{q^i} + \dot T^i_\alpha \partial_{v^i} & u_{1,\alpha} &= T^i_\alpha \partial_{v^i}\,.
\end{align}
Imposing that the action be invariant under variations with respect to $u_\alpha$, we get
\begin{equation}\label{eq:u var S}
	\Lie_{u_\alpha} S_1 = \int_\gamma \lf[ \iota_X \lf( \iota_{u_\alpha} \omega + \de J_\alpha \rt) - \lf(\iota_{u_\alpha} \de H\rt) \de t \rt] = 0\,,
\end{equation}
where we have defined the closed (though possibly degenerate) 2-form $\omega = \de \theta$ and the quantities
\begin{equation}
	J_\alpha := \iota_{u_\alpha} \theta\,.
\end{equation}

\subsubsection{Noether's first theorem}

We gain intuition about the $J_\alpha$'s by considering the simplified case where the $\epsilon_\alpha$ are constants so that only $u_{0,\alpha}$ contributes to $u_\alpha$. The exterior derivative $\de$ on $\mathcal{V}_t$ reduces to the exterior derivative $\de_\mathcal{V}$ on $\mathcal{V}$ so that, for arbitrary $X$, \eqref{eq:u var S} leads to
\begin{align}
	\iota_{u_{0\alpha}} \omega + \de_\mathcal{V} J_\alpha &= 0 \label{Mdef1} \\
	\iota_{u_{0\alpha}} \de_\mathcal{V} H = 0 \label{sympNT} \,.
\end{align}
The first equation, states that $J_\alpha = \iota_{u_{0,\alpha}} \theta = T^{i}_{\alpha} \,\diby{L}{\dot q^i} $ is the charge whose Hamilton vector field $u_{0,\alpha}$ generates the global symmetry. In Section \ref{MM}, we will see that \eqref{Mdef1} is equivalent to the statement that $J_\alpha$ is a \textit{moment} of a strongly Hamiltonian group action cf. \cite[p.42]{woodhouse:1997}. Since the $T^i_\alpha$ form represenations of the original algebra $\mathfrak g$, the fields $u_{0,\alpha}$ are representations of $\mathfrak g$. Since the  moments $J_\alpha$ collapse elements of $\mathfrak g$ to numbers, they form representation of the dual algebra $\mathfrak{g}^*$. The collection of moments then form the  \emph{momentum map} of the group $G$ about which we will say more later.

The second equation \eqref{sympNT} requires that the Hamiltonian flow of $J_\alpha$ be zero: $\dot J_\alpha := \pb{J_\alpha}{H} = 0$, where $\pb \cdot \cdot$ is the Poisson bracket associated with the (non-degenerate) symplectic 2-form $\omega$. This is the symplectic version of Noether's first theorem, implying that the moments $J_\alpha$ are conserved \emph{Noether charges} cf. \cite[4.5.11]{ortega:2013}.

\subsubsection{Recovering the canonical constraint formalism}

We return now to the case where the gauge parameter $\epsilon$ can be an arbitrary function of time. Inserting the prolongation \eqref{eq:prolongation} into the variation \eqref{eq:u var S} leads to several independent constraints. First, since each derivative of $\epsilon$ is an independent, freely specifiable function at a single time, one obtains two independent constraints: one for the $\epsilon$ term and one for the $\dot\epsilon$ term. These terms split further into 1-form equations on $\mathcal V$ and terms multiplying $\de t$. The net result is
\begin{align}\label{eq:simple cascade}
	\iota_{u_{i,\alpha}} \omega + \de_\mathcal{V} J_{i,\alpha} &= 0 \quad (i = 0,1) & \iota_{u_0,\alpha} \de_\mathcal{V} H &= 0 & \iota_{u_{1,\alpha}} \de_\mathcal{V} H &= J_0\,,
\end{align}
where $J_{i,\alpha} := \iota_{u_{i,\alpha}}\theta$ are the moments generating the $u_{i,\alpha}$.

Because $u_{1,\alpha}$ has no $\de q$ leg, $J_{1,\alpha} = \iota_{u_{1,\alpha}} \theta = 0$ is exactly zero. Thus, the $i=1$ component of the first equation becomes
\begin{equation}
	\iota_{u_{1,\alpha}} \omega = 0\,.
\end{equation}
This says that $u_{1,\alpha}$ is a degenerate direction of $\omega$. Using the explicit definitions \eqref{eq:ui} and \eqref{eq:theta} (recalling that $\omega = \de\theta$), we find that
\begin{equation}\label{eq:null W}
	T^i_\alpha W_{ij} = 0\,,
\end{equation}
which reproduces our earlier result of \eqref{Hessker} that the symmetry generators are in the kernel of the Hessian. Note that this is an off-shell identity, reflecting the non-invertibility of the Legendre transform. Since $W_{ij}$ reflects the part of the Hessian affecting the $v^i \to p_i$ transformation, it states that the vector $u_{1,\alpha} = T^i \partial_{v^i}$ projects to zero under the Legendre transform.

Using the explicit expression for $H$ from \eqref{eq:hamiltonian}, we find that the last equation of \eqref{eq:simple cascade} becomes
\begin{equation}
	J_{0,\alpha} = \iota_{u_{1,\alpha}} \de H = T^i_\alpha W_{ij} v^j = 0\,.
\end{equation}
This leads to the \emph{primary constraint}
\begin{equation}
	J_{0,\alpha} = T^i_\alpha \diby{L}{v^i} = T^i p_i = 0\,.
\end{equation}
This off-shell identity is linked to the non-invertibility of the Legendre transform apparent from \eqref{eq:null W}. Because of this, it can only be expressed in phase space and not velocity phase space. In this way, the primary constraint $J_{0,\alpha} = 0$ replaces the vector $u_1$ on phase space, which can only be expressed on velocity phase space. So while $J_{0,\alpha}$ reduces the phase space degrees of freedom by imposing a constraint, $u_1$ reduces the corresponding velocity phase space degrees of freedom by enlarging the kernel of $\omega$.

The primary constraint $J_{0,\alpha} = 0$ can be inserted into the $0$-component of the first equation of \eqref{eq:simple cascade} to imply that $u_{0,\alpha}$ is another null direction of $\omega$ off-shell:
\begin{equation}
	\iota_{u_{0,\alpha}} \omega = 0\,.
\end{equation}
On phase space, this null vector is a consequence of restricting $\omega$ to the constraint surface $J_{0,\alpha} = 0$. This condition can be combined with the second equation of \eqref{eq:simple cascade} and simplified using \eqref{eq:null W} to give the standard \emph{Lagrangian constraint}:
\begin{equation}
	T^{i}_{\alpha} K_i = 0\,,
\end{equation}
where $K_i$ is given by \eqref{kequation}. This calculation involves eliminating the $\dot T^i_\alpha$ term appearing in both equations.

The central result of the preceding analysis is that our system of equations resulting from the symmetries of $S_1$ leads to: i) an initial value constraint $J_{0,\alpha} = 0$ on phase space; ii) a null direction $u_{0,\alpha}$ of $\omega$ associated with this constraint and; iii) a second null direction $u_{1,\alpha}$ replacing the role of $J_{0,\alpha}$ on velocity phase space. These results can be used to derive the Lagrangian constraints and kernel of the Hessian. Together they provide a geometric framework that reproduce the standard results from constrained Hamiltonian theories in the presence of purely first class primary constraints but within a Noether rather than Dirac style analysis. 

A final significant formal point runs as follows. By construction $u_{\alpha}$ is the linear combination of the $u_{i,\alpha}$ which reproduces the original Lagrangian symmetries when projected on to the configuration variables $q^i$. Because $u_{1,\alpha}$ has no $\de q$ leg, the symmetry is generated entirely by $u_{0,\alpha}$. On phase space, with the Poisson bracket canonically extended off the constraint surface, the first equation of \eqref{eq:simple cascade} says that $u_{0,\alpha}$ is the Hamilton vector field of $J_{0,\alpha}$. Thus, the gauge generator is simply the primary constraint $J_{0,\alpha}$ itself, reproducing Dirac's theorem of Section~\ref{sec:Dirac thm}.

To get a well-defined evolution on phase space, one then needs to both restrict the evolution to the primary constraint surface defined by $J_{0,\alpha}^{-1}(0) = \{\, m \in \Gamma : J_{0,\alpha}(m) = 0 \ \text{ for all } \alpha \,\}$ then quotient by the action of $u_{0,\alpha}$. To find the Lie algebra obeyed by $u_{0,\alpha}$, note that prolongation is a Lie algebra homomorphism as outlined in \cite[Prop. 5.15]{Olver:2000}. The full prolongations $u_\alpha$ then form representations of the original algebra $\mathfrak g$:
\begin{equation}
	[u_\alpha, u_\beta] = f^\gamma_{\alpha\beta} u_\gamma\,.
\end{equation}
Using \eqref{eq:prolongation} and collecting terms with the same time derivatives of $\epsilon$ gives
\begin{align}\label{eq:ui alg}
	[u_{0,\alpha}, u_{0,\beta}] &= f^\gamma_{\alpha\beta} u_{0,\gamma} &
	[u_{0,\alpha}, u_{1,\beta}] &= f^\gamma_{\alpha\beta} u_{1,\gamma} &
	[u_{1,\alpha}, u_{1,\beta}] &= 0\,.
\end{align}
On phase space, where $u_{1,\alpha}$ is projected out, this reduces to the original algebra $\mathfrak g$ of the symmetry group on configuration space. Thus, a well-defined evolution can be achieved on phase space by additionally quotienting by this group:
\begin{equation}\label{redphase}
	\Gamma_\text{red} = J_{0,\alpha}^{-1}(0)/G\,.
\end{equation}
On velocity phase space, however, there is no primary constraint. Instead, the kernel of $\omega$ must include $u_{1,\alpha}$.\footnote{Note that, on phase space, the symplectic structure is normally canonically extended off the primary constraint surface instead.} The reduced space can be obtained by quotienting by the full graded algebra $\mathfrak g_\text{ext} = TG_\text{ext}$ given in \eqref{eq:ui alg}
\begin{equation}
	\mathcal V_\text{red} = \mathcal V / G_\text{ext}\,.
\end{equation}
The focus of the following final section is to consider two approaches for applying the tools of category theory to the to the construction of the reduced phase space as defined by \eqref{redphase}. In essence, the contrast will be to understand $\Gamma_\text{red}$ as either an object in the codomain of a functor following \cite{bradley:2025}  or as the composite of a special class of arrows following \cite{landsman:2001,landsman:2005}. 
 
\section{Reduction and Nomic Regularity}
\label{Reduction and Nomic Regularity}
\subsection{Symplectic Reduction and Category Theory}
\label{Bradley}

Let us return to the intrinsic presentation of a classical constrained Hamiltonian theory and consider the procedure of \textit{symplectic reduction}. We will initially follow the insightful analysis of \cite{bradley:2025} who proceeds in three steps. First, assume that there exists a smooth, differentiable manifold, the reduced phase space, $\Gamma_R$, defined by taking the quotient of $\Sigma$ by the kernel of $\tilde{\omega}$.\footnote{See \cite[p.212]{ortega:2013} for full formal details.} Next, define an open, surjective projection map $\pi : \Sigma \to \Gamma_R$ such that we define the reduced two-form $\omega_R$ via
$\tilde{\omega} = \pi^{*}(\omega_R)$, which acts according to $\omega_R(X_{A_R},X_{B_R}) = \tilde{\omega}(X_{\tilde{A}},X_{\tilde{B}})$ where
$X_{A_R} = \pi_{*}(X_{\tilde{B}})$ and $X_{B_R} = \pi_{*}(X_{\tilde{B}})$. When the map $\pi$ is well-defined there can expect $\omega_R$ to be a symplectic two form.\footnote{See \cite[p.212]{ortega:2013} for full formal details.} Finally, one can then define a reduced Hamiltonian $H_R$ via $\pi_*(\tilde{H})$ and formulate Hamilton's equations on the reduced phase space as $\omega_R(X_{H_R})=dH_R$. The theory formulated on reduced phase space has regular nomic structure meaning that it provides us with a well-posed initial value problem that avoids the pernicious redundancy we encountered earlier. This matches precisely the conclusion of \cite{bradley:2025}, that removing this form of redundancy is well-motivated from the perspective of pursuing a well-posed initial value problem (p. 14). 

The next step is to  provide a category theoretic  presentation of the reduced theory via the category \textbf{HamRed} which has objects $(\Gamma_R,\omega_R,H_R)$ and arrows between
objects $(\Gamma^1_R,\omega^1_R,H_R^1)$ and $(\Gamma^2_R,\omega^2_R,H_R^2)$ given by diffeomorphisms $h : \Sigma^1_R \to\Sigma^2_R$ such that $h^{*}(\omega^2_R) = \omega^1_R$ and $h^{*}H_R^2 = H_R^1$. The category \textbf{HamRed} has an elegant simplicity since models are picked out by symplectic geometries and arrows by the symplectomophisms. In order to study 
the relationship between the intrinsic presentation the classical constrained Hamiltonian theory and the theory of the reduced phase space \cite{bradley:2025} then defines the functor $G$ that takes an object $(\Sigma,\tilde{\omega},\tilde{H})$ to $(\Gamma_R,\omega_R,H_R)$ and that takes an arrow $g : \Sigma \to \Sigma$ to $g_R : \Gamma_R\to \Gamma_R$. Crucially, since the reduction map corresponds to a projection of equivalence classes defined by directions on $\Sigma$ defined by the kernel of $\tilde{\omega}$, for any $g$ that acts by moving along those directions the corresponding $g_R$ will be the identity map. This allows \cite{bradley:2025} to prove that the functor $G$ is full and essentially surjective but not faithful (see her Proposition 4). In our terminology we have that the intrinsic formulation of the theory of the constraint surface of the original phase space has surplus representational capacity compared to the theory on the reduced phase space relative to the functor defined by the reduction map. Equivalently there are distinct symmetry related DPMs in unreduced formalism that get mapped to the same DPMs in the reduced formalism. 

The fact that the functor is full and essentially surjective means that symplectic reduction does not remove \textit{further} surplus structure compared to the intrinsic presentation of the theory on the constraint surface. This means one could plausibly understand the passage from a classical constrained Hamiltonian theory formulated extrinsically on the phase space to a reduced theory in terms of a pair of functors defined via: 1) the immersion map that takes us from $\textbf{ConHam}$ to $\textbf{PreHam}$ and removes (only) surplus structure by enlarging the isomorphism class; and 2) the quiotenting map that takes us from $\textbf{ConHam}$ to $\textbf{HamRed}$ and removes (only) surplus representational capacity by projecting non-trivial to trivial automorphisms. Since the composition of two functors is always a functor there should then be definable a functorial relationship between the unreduced and reduced theory. This functor would be expected to be essentially surjective but neither full nor faithful. 

The `symplectic reduction as functor' perspective is not, however, the approach taken by mathematical physicists in formalising the symplectic reduction in terms of category theory. There are four interconnected issues that motivate an alternative formalisation of symplectic reduction to that provided by \cite{bradley:2025}. First, it would be helpful to have available a more unified formalisation of the immersion and quotient maps. Second, the reduction procedure as we defined it does not make explicit the conditions for the reduced phase space to inherent manifold structure and avoid singularities. Third, it would be more natural for the initial `input' of the reduction procedure to, like the output, be given simply by a symplectic manifold and for us to have available a category theoretic analysis which encodes this. Fourth, and ultimately, most crucial, moving to an approach in which state spaces are understood as arrows and reduction as arrow composition allows for direct connection with procedures for the quantization of theories with irregular nomic structure that was Dirac's principal motivation in constructing the constrained Hamiltonian formalism in the first place. In order to understand the structure of this alternative category theoretic presentation of symplectic reduction we must return to the idea of the momentum map associated with a symmetry which was mentioned above. 

\subsection{Poisson Manifolds and the Momentum Map}
\label{MM}

Following \cite{landsman:2001,landsman:2005} cf. \cite{ButterfieldPP,ortega:2013}, consider the generalisation of symplectic manifolds to the class of \textit{Poisson manifolds}, $M$, which, by definition, are equipped with a Lie bracket that acts as $\{\cdot,\cdot\}:C^\infty(M) \times C^\infty(M) \rightarrow C^\infty(M)$ and is such that for each $A\in C^\infty(M)$ the map $B \rightarrow \{A,B\}$ defines a derivation of $C^\infty(M)$ which can be identified with a Hamilton vector field $X_f$. A smooth map between Poisson manifolds is then a \textit{Poisson map} when its pullback is a Lie algebra homomorphism and is a \textit{anti-Poisson map}  when its pullback is a Lie algebra anti-homomorphism. Note that the space containing a single point $\text{pt}$ is a Poisson manifold with trivial Lie bracket and for any Poisson manifold $M$ the map $M\rightarrow \text{pt}$ is trivially both Poisson and anti-Poisson.

Now, consider a Lie group $G$ with associated Lie algebra $\mathfrak{g}$ which has a \textit{strongly Hamiltonian action} on $M$.  This implies that there exists a pair of Lie algebra homomorphisms: i) $x\in \mathfrak{g} \mapsto X^M \in \Gamma(M,TM)$; and ii) $x\in \mathfrak{g} \mapsto J_x \in C^\infty(M)$ with the property that $X^M = X_{J_{x}}$, where $\Gamma(M,TM)$ indicates the space of all smooth vector fields on $M$ and $J_x$ are the special functions who generate Hamilton vector field $X^M_{J_{x}}$ corresponding to each Lie algebra element $x$. The dual vector space to $\mathfrak{g}$ then defines a Poisson manifold which we indicate as $\mathfrak{g}^*$. Finally, we can then understand the collection of the functions $J_x$ to define the momentum map of $\mathfrak{g}$ as the Poisson map:
\begin{equation}
J: M \rightarrow \mathfrak{g}^*
\end{equation}    
defined via the point-wise condition that for all $m\in M$ and $x\in \mathfrak{g}$ we have that $\langle J(m),x\rangle=J_x(m)$ and $\langle\cdot,\cdot\rangle: \mathfrak{g}^* \times \mathfrak{g} \rightarrow \mathbb{R}$ is the duality pairing. Connection with our earlier analysis can be straightforwardly seen by considering orthonormal bases $e^\alpha$ in the algebra $\mathfrak g$ and $\tilde e^\alpha$ in the dual algebra $\mathfrak g^*$. Using these bases, we can identify $J_x = J_{0,\alpha} \tilde e^\alpha$ and $X^M = u_{0,\alpha} e^{\alpha}$. Then, as we saw in Section~\ref{sec:VPS construction}, Equation~\ref{eq:simple cascade} confirms that $X^M$ is the Hamilton vector field generated by $J_x$ and that these quantities are valued in the appropriate representations of $\mathfrak g$.
 
 Physical intuition for the rather abstract momentum map formalism can be gained immediately by the recognition that the condition $X_{J_X} (H)=0, \: \forall x\in\mathfrak{g}$ implies that the $J_x$ are the constants of the motion associated with the elements of the group $G$ via the symplectic version of Noether's first theorem. Thus, the momentum map for translations defines the linear momentum at every point and the momentum map for rotations defines the angular momentum at every point. See \cite[\S6.3]{ButterfieldPP} for explicit details. 
 
The power and elegance of the momentum map formalism is made immediately apparent by the representation of symplectic reduction that it affords. Let us consider a symplectic manifold $(\Gamma,\omega)$ acted on by a Lie group $G$ with associated Lie algebra $\mathfrak{g}$ that is strongly Hamiltonian action on $\Gamma$. The associated constraint surface is then defined simply by the condition introduced above that $J^{-1}(0)=\{\,m\in M : J_X(m)=0 \text{ for all } X\in\mathfrak g\,\}$ and the reduced phase space via:
\begin{equation}
\Gamma_R= J^{-1}(0)/G
\end{equation}
The immersion map and the quotient map as can then be expressed as:
\begin{equation}
i:J^{-1}(0) \hookrightarrow \Gamma \: \: \: \:  \pi:  J^{-1}(0) \rightarrow \Gamma_R
 \end{equation}
In the case that $0$ is a regular value of $J$ and the $G$ action is proper and free on $J^{-1}(0)$ we are guaranteed that $\Gamma_R$ is a manifold with symplectic structure which is such that $i^* \omega = \pi^* \omega_R$. This defines the Marsden-Weinstein reduction under which  starting from a symplectic manifold $(\Gamma,\omega)$ equipped with a strongly Hamiltonian group action $G$ we produce a new symplectic manifold $(\Gamma_R,\omega_R)$ in which the group action has been `quotiented out'.\footnote{This process of immersion and reduction is precisely the picture that emerges from the considerations of Section~\ref{sec:VPS construction} on phase space. Note that, on \emph{velocity} phase space, the picture is different. Rather than starting with an immersion onto the primary constraint surface and then quotienting by the gauge group $G$, you must quotient by the full action $G_\text{ext}$ of the extended gauge group formed by the full set $u_{i,\alpha}$.} So formulated, however, it is clear that we should \textit{not} understand the process of symplectic reduction in terms of a pair of functors that take us from the category of symplectic manifolds to itself. \textit{The immersion map is not total since it is not defined for arbitrary symplectomorphism}. In fact, once defined in terms of momentum maps the natural category theoretic presentation of symplectic reduction is in terms of arrow composition. The following two sections are devoted to the explication and interpretation of this idea following the account of \cite{landsman:2001,landsman:2005}.

\subsection{State Spaces as Arrows}

The formal basis for presenting the state space of a constrained Hamiltonian theory as an arrow within a category rather than an object comes from the notion of a \textit{symplectic dual pair}. In general, a symplectic dual pair is defined by a symplectic manifold $(\Gamma,\omega)$, a pair of Poisson manifolds, $P,Q$, and a pair of maps given by the anti-Poisson map $\Gamma\rightarrow P$ and the Poisson map $Q\leftarrow \Gamma$.  We require that the pullback of any function on $P$ should Poisson commute on $\Gamma$ with the pullback of any function on $Q$. We write the symplectic dual pair itself as:
\begin{equation}
Q \xleftarrow{q} \Gamma  \xrightarrow{p} P
\end{equation} 
and thus can write the pullback condition as $\{p^*f,q^*g\}= 0$ for all $f\in C^\infty(P)$ and $g\in C^\infty(Q)$. 

Formally, a symplectic dual pair is an example of a \textit{bimodule}. Bimodules can be understood as a generalization of homomorphisms since we can construct a functor that takes us from the category that has algebras as objects and homomorphisms as arrows to a category which has $k$-algebras as objects, and bimodules as arrows \citep[2.1]{landsman:2001}, cf. \cite{feintzeig:2024}. Bimodules always come with two compatible `left' and `right' actions. In the case of a bimodule over a pair of Poisson manifolds $P,Q$ as defined by a symplectic dual pair these are precisely the anti-Poisson (left) and Poisson (right) maps from the symplectic manifold to $P$ and $Q$ respectively. 

The condition for two symplectic dual pairs over two Poisson manifolds $P,Q$ to be isomorphic is then that there exists a symplectomorphism renders the relevant Poisson and anti-Poisson maps identical. That is, for the dual pairs:
\begin{eqnarray}
Q \xleftarrow{q_1} \Gamma_1  \xrightarrow{p_1} P \\
Q \xleftarrow{q_2} \Gamma_2  \xrightarrow{p_2} P
\end{eqnarray}  
the condition for isomorphism is that we have that $\varphi:\Gamma_1\rightarrow \Gamma_2$ is such that $q_2\varphi = q_1$ and $p_2\varphi = p_1$. Following \cite{landsman:2005}, the equivalence class of isomorphic symplectic dual pairs can then be interpreted as an arrow from $Q$ to $P$. So defined, the arrows are well-suited to the representation of model spaces of both constrained and unconstrained Hamiltonian theories. 

Consider for example, the simplest case of a theory with regular nomic structure. The phase space of such a theory can be represented simply via a symplectic manifold and we can choose as the appropriate dual pair simply the trivial Poisson manifolds and maps given by:
\begin{equation}
\textit{pt} \leftarrow (\Gamma,\omega)  \rightarrow \textit{pt}
\end{equation} 
where \textit{pt} indicates the one-point manifold that is trivially both Poisson and anti-Poisson. Since we define the arrows as the equivalence class of isomorphic symplectic dual pairs the condition for two model spaces as arrows to be isomorphic is identical to that which we would obtain by considering model spaces to be objects in the category of symplectic manifolds with symplectomorphisms again the salient isomorphisms. Thus if we take as a norm for physical representational practice that mathematical objects which are isomorphic under the relevant mathematical notion of isomorphism should be taken to have the same representational capacities \citep{weatherall:2018} then we immediately get that model spaces as symplectic manifolds have the same representational capacities whether we take these to be objects in the category of symplectic manifolds \citep{Thebault2026} or dual pairs as arrows between (trivial) Poisson manifolds.\footnote{The generalisation of this approach to richer representations that include dynamics as encoded within non-trivial Poisson structure should then follow by considering the \textit{bifibrations} defined respectively by the orbits of a Hamiltonian vector field $X_H$ and the (connected components of the) level sets of $H$ \citep{fasso:2005}.}
 
The dual pair representation of model spaces as arrows is particularly well-suited to the representation of constrained Hamiltonian theories. As noted already, the dual to a Lie algebra, $\mathfrak{g}^*$, is a Poisson manifold. This means we can consider Poisson maps defined via $\mathfrak{g}^*\leftarrow (\Gamma,\omega) $ and anti-Poisson map $ (\Gamma,\omega) \rightarrow \mathfrak{g}^*_{-}$ where $\mathfrak{g}^*_{-}$ is the relevant Poisson manifold equipped with minus the Poisson bracket. Now consider a constrained Hamiltonian system with unconstrained phase space $(\Gamma,\omega)$ and constraint surface defined via  $J=0$ where $J:\Gamma \rightarrow \mathfrak{g}^*$ is the momentum map associated to strongly Hamiltonian action of a Lie group $G$ on $\Gamma$. Following \cite[\S8]{weinstein:1983}, in general we have that if a Lie group $G$ acts freely on a symplectic manifold $\Gamma$ with momentum map $J$, then the  space $\Gamma/G$ is a manifold and functions on $\Gamma/G$ indicates indicates quotient for a right $G$-action and may thus be identified with the function group of $G$-invariant functions. 

We thus have that the manifold $\Gamma/G$ has Poisson structure for which the projection $\rho: \Gamma \rightarrow \Gamma/G$ is a Poisson mapping. Furthermore, the structure of the maps $\rho$ and $J$ is such that their pull-backs to $\Gamma$ will Poisson commute. Thus the model space of a constrained Hamiltonian theory can be defined via symplectic dual pair as \citep{weinstein:1983,landsman:2005}:
\begin{equation}
G\backslash \Gamma \xleftarrow{\rho} (\Gamma,\omega) \xrightarrow{J} \mathfrak{g}^*_{-}
\end{equation} 
where $G\backslash \Gamma$ indicates quotient for a left $G$-action.  In fact, one of the motivating examples in the construction of the more general class of mathematical object defined by dual pairs was precisely the pair above obtained by considering a strongly Hamiltonian Lie group action on a symplectic manifold with associated momentum map. It is thus not surprising that one provides a natural way to represents the other. 

The proposal of \cite{landsman:2001,landsman:2005} is then to treat the equivalence class of isomorphic symplectic dual pairs as arrows. This then amounts to interpreting the model space of a constrained Hamiltonian theory together with the associated maps, $\rho$ and $J$, as defining arrows between the Poisson manifolds $G\backslash \Gamma$ and  $\mathfrak{g}^*_{-}$. Since such arrows are defined directly via the equivalence class we automatically get that symplectomorphic dual pairs correspond to model spaces with equivalent representational capacity. An attractive feature of this approach is that the extra structure defined by the left and right `legs' of the dual pair allows us to encode the partition and projection that pick out the restrictions required for the models to be respectively dynamically possible and distinct. These are the immersion map $i:J^{-1}(0) \hookrightarrow \Gamma$ and quotient map $ \pi:  J^{-1}(0) \rightarrow \Gamma_R$ each of which is defined via the momentum map. 

The symplectomorphisms that define the equivalence class of a dual pair are then required to be such that if $\varphi:\Gamma \rightarrow \Gamma'$ is the symplectomorphism then $J'\varphi = J$. Evidently, the momentum map encodes nomic structure in terms of the partition and projection functions. Such nomic structure is automatically defined only up to isomorphism and can be specified within the same object that defines the state space up to isomorphism. Constrained Hamiltonian theories can then be understood in terms of a collection of arrows and we automatically recover the basic requirement that theories with isomorphic state spaces will be equivalent. The next step is cateorgical presentation is to specify the composition operation. The following subsection will introduce the appropriate operation and consider the representation of symplectic reduction as special class of arrow compositions under this product.  

\subsection{Symplectic Reduction as Arrow Composition}

Let us consider two compatible dual pairs $Q \leftarrow \Gamma_1 \to P$ and $P \leftarrow \Gamma_2 \to R$ and the conditions for which we can define a tensor product that yields a new dual pair $Q \leftarrow \Gamma_3 \to R$. Let us consider a symplectic manifold $(\Gamma,\omega)$ and let $\Sigma$ be a closed sub-manifold. As above we define the closed two form $\tilde{\omega}=i^*\omega$ as the restriction of $\omega$ to $\Sigma$ via $i:\Sigma \hookrightarrow M$. 

Define the kernel of $\tilde{\omega}$ as the \text{null distribution} on $\Sigma$ which we will write as $\mathcal{N}_\Sigma$. Following \cite{landsman:2001}, when the rank of $\tilde{\omega}$ is constant on $\Sigma$ the null distribution $\mathcal{N}_\Sigma$ is
smooth and completely integrable. Let us denote the corresponding foliation $\mathcal{F}_\Sigma$ and assume that the space $\Sigma/\mathcal{F}_\Sigma$ of leaves of this foliation is a manifold in its natural topology. These assumptions suffice to restrict to cases in which symplectic reduction is non-singular and we are guaranteed that the reduced space will be a symplectic manifold. 

Under these restrictions we can define the product of dual pairs as follows. Let $Q \leftarrow \Gamma_1 \to P$ and $P \leftarrow \Gamma_2 \to R$ be dual pairs, with Poisson maps $J_L : M_1 \to P^{-}$ and $J_R : M_2 \to P$. Assume that
\begin{equation}
T_pP = (T_x J_L)(T_x \Gamma_1) \oplus (T_y J_R)(T_y \Gamma_2)
\end{equation}
for all $(x,y) \in \Gamma_1 \times_P \Gamma_2$, where $p = J_L(x) = J_R(y)$. Define the fibre product between two manifolds by $ M_1 \times^{f,g}_P M_2 = \{(m_1,m_2)\in M_1 \times M_2 | f(m_1)=g(m_2)\}$ for $f:m_1\rightarrow P$ and $g:m_2\rightarrow P$. Then we can specify that $\Gamma_3 = \Gamma_1 \times \Gamma_2$ and $\Sigma= \Gamma_1 \times_P \Gamma_2$ and are guaranteed that $\mathcal{N}_\Sigma$ is
smooth and completely integrable. Given that $\Sigma/\mathcal{F}_\Sigma$ is a manifold in its natural topology we then have that the product operation $\Gamma_1 \circledcirc_{P} \Gamma_2$ can be defined via the symplectic manifold:
\begin{equation}
\Gamma_1 \circledcirc_{P} \Gamma_2 = (\Gamma_1 \times_P \Gamma_2)/N_\Sigma. 
\end{equation}
and that 
\begin{equation}
Q \leftarrow \Gamma_1 \circledcirc_{P} \Gamma_2 \to R.
\end{equation}
defines a dual pair. So defined the composition operation is associative up to isomorphism. 

If we consider the case of a connected Lie group $G$ then the product of the dual pairs defined by a constrained Hamiltonian theory  $G\backslash \Gamma \xleftarrow{\rho} (\Gamma,\omega) \xrightarrow{J} \mathfrak{g}^*_{-}$ and  the zero coadjoint orbit $\mathfrak{g}^*_{-} \hookleftarrow 0 \to \mathrm{pt}$ gives: 
\begin{equation}
G\backslash \Gamma \xleftarrow{\rho} (\Gamma,\omega) \xrightarrow{J} \mathfrak{g}^*_{-}  \circledcirc_{\mathfrak{g}^*_{-}}\mathfrak{g}^*_{-} \hookleftarrow 0 \to \mathrm{pt} \cong G \backslash \Gamma \leftarrow \Gamma_R \to \mathrm{pt}
\end{equation}
where $\Gamma_R = J^{-1}(0)/G$ is the Marsden--Weinstein quotient. This reconstruction of the Marsden--Weinstein reduction procedure motivates \cite{landsman:2005} to introduce the category \textbf{Poisson} whose objects are regular Poisson manifolds and arrows are equivalence classes of regular symplectic dual pairs. Under this restriction we get that all products exist and there are identity arrows \citep[Def. 5.10]{landsman:2001}. Whilst the regularity condition is relatively mild in the case of the Poisson manifolds it is restrictive enough in the case of the dual pairs to exclude  $\mathrm{pt}\hookleftarrow \Gamma_R \to \mathfrak{g}^*_{-}$. This motivates the reconstruction of symplectic reduction as the well defined composition:

\begin{equation}
 \mathrm{pt} \leftarrow (\Gamma,\omega) \xrightarrow{J} \mathfrak{g}^*_{-}  \circledcirc_{\mathfrak{g}^*_{-}}\mathfrak{g}^*_{-} \leftarrow 0 \to \mathrm{pt} \cong \mathrm{pt} \leftarrow \Gamma_R \to \mathrm{pt}
\end{equation}
Since arrows are equivalence classes of regular symplectic dual pairs we have the arrow that is the output of the reduction procedure $\mathrm{pt} \leftarrow \Gamma_R \to \mathrm{pt}$ is defined only up to symplectomorphism. 

If we then define the irregular representations of a constrained Hamiltonian theories by collections of arrows of the form $\mathrm{pt} \leftarrow (\Gamma,\omega) \xrightarrow{J} \mathfrak{g}^*_{-}$ and reduction by composition with the zero coadjoint orbit $\mathfrak{g}^*_{-} \leftarrow 0 \to \mathrm{pt}$ then the collection of arrows $\mathrm{pt} \leftarrow \Gamma_R \to \mathrm{pt}$ correspond to the \textit{regular representations} of the theory. That is, representations in which the pernicious form of surplus representational capacity has been eliminated and the corresponding initial value problem rendered well-posed. This suggests a notion of theoretical equivalence at the level of regular representations: two constrained Hamiltonian theories have equivalent regular representations when the reduced phase spaces are symplectomorphic. In essence, this means that two theories can be understood to be equivalent when stated as well-posed initial value problems on a reduced phase space but inequivalent when stated as ill-posed initial value problems as constrained Hamiltonian theories that include surplus representational capacity of the particular pernicious form. For example, we may have two theories, $T$ and $T'$, with the same underlying well-posed degrees of freedom but distinct constrained Hamiltonian representations as encoded in the momentum maps $J$ and $J'$ and group actions $G$ and $G'$ whenever we have that:
\begin{equation}
 J^{-1}(0)/G \cong J'^{-1}(0)/G'
\end{equation}
The symplectic reduction as arrow composition thus allows for a  category theoretic presentation of constrained Hamiltonian theories where we automatically gain an equivalence in regular representations based upon the definition of arrows as symplectic dual pairs up to isomorphism.


\section{Prospectus}

The foregoing analysis has articulated a range formal and interpretational features particular to theories with irregular nomic structure. For an important subclass of such theories, we have demonstrated the interconnections between pernicious underdetermination problems and well-posedness of the initial value formulation making use of a first order velocity-phase space formalism. This analysis has been shown to underpin the canonical constraint analysis in both its original form due to Dirac and in geometric form as per Marsden-Weinstein symplectic reduction. Two natural lines of generalisation of the classical analysis are to consider the cases of  irregular nomic structure which arises from non-isochronous symmetries and/or which involves secondary as well as primary constraints. The first generalisation would allow our analysis to include mechanical and field theories which are reparametrization invariant and would thus lead us into consideration of the problem of time. Work along these lines can be found in \cite[\S12]{gryb:2024}. The second generalisation  would allow our analysis to include theories such as electromagnetism in which the primary constraints do not automatically propagate. This analysis is provided in formal terms in \cite{Gryb:2026}. The intersection of these two generalisations, that is non-isochronous symmetries and secondary constraints, is precisely the case of general relativity. The application of the velocity-phase space analysis to this case is an outstanding mathematical problem.  

A third dimension of generalisation of the foregoing analysis is the extension to quantum theories. Ultimately, the principal motivation for studying the canonical formulation of irregular systems is the challenge of quantization. This is because standard techniques for the conversion of a classical to a quantum theory are built upon (in different senses) promotion of core parts of the regular nomic structure of a classical theory to corresponding quantum structures. Thus, we find that one route to the quantization of a theory with irregular nomic structure is precisely to `reduce first' and `quantize' second. There is, however, a second route, to `quantize first and `reduce second'. This is the procedure of constraint quantization or Dirac quantization. Furthermore, there exists a conjecture, proved in a wide variety of cases that the quantum theories produced via the two routes are equivalent. This is the famous Guillemin–Sternberg conjecture often informally expressed in the idea that `reduction commutes with quantization'. In a companion paper we will deploy the formal tools and interpretational concepts introduced in the present paper towards the analysis of constrain quantization and Guillemin–Sternberg conjecture.  Our goal will be to understand the relation between quantum and classical reduction procedures and evaluate the implications of the proposal that the functionality of quantization can be understood, in a certain sense, to be equivalent to the Guillemin–Sternberg conjecture.\footnote{This is specifically intended to allow connection with the work of \cite{feintzeig:2024b,feintzeig:2025} which applies the tools of category theory to quantization in the regular case.}

\section*{Acknowledgements}

We are grateful to Clara Bradley, Rami Jreige, and Jer Steeger for helpful discussion.

\bibliographystyle{chicago}
\bibliography{Masterbib3}

\begin{thebibliography}{}

\bibitem[\protect\citeauthoryear{Anderson}{Anderson}{2017}]{anderson:2017}
Anderson, E. (2017).
\newblock {\em The Problem of Time}.
\newblock Springer.

\bibitem[\protect\citeauthoryear{Anderson and Bergmann}{Anderson and
  Bergmann}{1951}]{anderson:1951}
Anderson, J.~L. and P.~G. Bergmann (1951).
\newblock Constraints in covariant field theories.
\newblock {\em Physical Review\/}~{\em 83\/}(5), 1018.

\bibitem[\protect\citeauthoryear{Baez and Shulman}{Baez and
  Shulman}{2009}]{baez:2009}
Baez, J.~C. and M.~Shulman (2009).
\newblock Lectures on n-categories and cohomology.
\newblock In {\em Towards higher categories}, pp.\  1--68. Springer.

\bibitem[\protect\citeauthoryear{{Barbour} and {Foster}}{{Barbour} and
  {Foster}}{2008}]{Barbour:2008}
{Barbour}, J. and B.~Z. {Foster} (2008, August).
\newblock {Constraints and gauge transformations: Dirac's theorem is not always
  valid}.
\newblock {\em ArXiv e-prints\/}.

\bibitem[\protect\citeauthoryear{Barbour and Bertotti}{Barbour and
  Bertotti}{1982}]{barbour:1982}
Barbour, J.~B. and B.~Bertotti (1982).
\newblock Mach's principle and the structure of dynamical theories.
\newblock {\em Proceedings of the Royal Society of London. A. Mathematical and
  Physical Sciences\/}~{\em 382\/}(1783), 295--306.

\bibitem[\protect\citeauthoryear{Barrett}{Barrett}{2020}]{Barrett2020b}
Barrett, T.~W. (2020).
\newblock Structure and equivalence.
\newblock {\em Philosophy of Science\/}~{\em 87\/}(5), 1184--1196.

\bibitem[\protect\citeauthoryear{Barrett}{Barrett}{2022}]{Barrett2020a}
Barrett, T.~W. (2022).
\newblock How to count structure.
\newblock {\em No\^{u}s\/}~{\em 56\/}(2), 295--322.

\bibitem[\protect\citeauthoryear{Belot}{Belot}{2003}]{Belot:2003}
Belot, G. (2003).
\newblock Symmetry and gauge freedom.
\newblock {\em Studies In History and Philosophy of Modern Physics\/}~{\em 34},
  189--225.

\bibitem[\protect\citeauthoryear{Belot}{Belot}{2013}]{Belot:2013}
Belot, G. (2013).
\newblock Symmetry and equivalence.
\newblock In R.~Batterman (Ed.), {\em The Oxford Handbook of Philosophy of
  Physics}. Oxford University Press.

\bibitem[\protect\citeauthoryear{Bradley}{Bradley}{2025a}]{Bradley2025b}
Bradley, C. (2025a).
\newblock Do first-class constraints generate gauge transformations? a
  geometric resolution.
\newblock {\em The British Journal for the Philosophy of Science\/}.
\newblock Published online 12 November 2025.

\bibitem[\protect\citeauthoryear{Bradley}{Bradley}{2025b}]{bradley:2025}
Bradley, C. (2025b).
\newblock Excess structure in the constrained hamiltonian formalism.
\newblock {\em Philosophy of Science\/}, 1--19.

\bibitem[\protect\citeauthoryear{Bradley}{Bradley}{2025c}]{Bradley2025a}
Bradley, C. (2025c).
\newblock Excess structure in the constrained hamiltonian formalism.
\newblock {\em Philosophy of Science\/}.
\newblock Published online 24 November 2025.

\bibitem[\protect\citeauthoryear{Bradley}{Bradley}{2025d}]{Bradley2025c}
Bradley, C. (2025d).
\newblock The relationship between lagrangian and hamiltonian mechanics: The
  irregular case.
\newblock {\em Philosophy of Physics\/}.
\newblock Published online 17 July 2025.

\bibitem[\protect\citeauthoryear{Bradley and Weatherall}{Bradley and
  Weatherall}{2020}]{bradley:2020}
Bradley, C. and J.~O. Weatherall (2020).
\newblock On representational redundancy, surplus structure, and the hole
  argument.
\newblock {\em Foundations of Physics\/}~{\em 50\/}(4), 270--293.

\bibitem[\protect\citeauthoryear{Butterfield}{Butterfield}{2007}]{ButterfieldPP}
Butterfield, J. (2007).
\newblock On symplectic reduction in classical mechanics.
\newblock In J.~Butterfield and J.~Earman (Eds.), {\em Philosophy of Physics},
  pp.\  1--131. Elsevier.

\bibitem[\protect\citeauthoryear{Casadio, Chataignier, Kamenshchik, Pedro,
  Tronconi, and Venturi}{Casadio et~al.}{2024}]{CASADIO2024169783}
Casadio, R., L.~Chataignier, A.~Y. Kamenshchik, F.~G. Pedro, A.~Tronconi, and
  G.~Venturi (2024).
\newblock Relaxation of first-class constraints and the quantization of gauge
  theories: From ``matter without matter'' to the reappearance of time in
  quantum gravity.
\newblock {\em Annals of Physics\/}~{\em 470}, 169783.

\bibitem[\protect\citeauthoryear{Dewar}{Dewar}{2022}]{Dewar2022}
Dewar, N. (2022).
\newblock {\em Structure and Equivalence}.
\newblock Cambridge University Press.

\bibitem[\protect\citeauthoryear{D{\'\i}az, Higuita, and Montesinos}{D{\'\i}az
  et~al.}{2014}]{diaz:2014}
D{\'\i}az, B., D.~Higuita, and M.~Montesinos (2014).
\newblock Lagrangian approach to the physical degree of freedom count.
\newblock {\em Journal of Mathematical Physics\/}~{\em 55\/}(12), 122901.

\bibitem[\protect\citeauthoryear{Diaz and Montesinos}{Diaz and
  Montesinos}{2018}]{diaz:2018}
Diaz, B. and M.~Montesinos (2018).
\newblock Geometric lagrangian approach to the physical degree of freedom count
  in field theory.
\newblock {\em Journal of Mathematical Physics\/}~{\em 59\/}(5), 052901.

\bibitem[\protect\citeauthoryear{D{\'\i}ez, Maier, M{\'e}ndez-Zavaleta, and
  Tehrani}{D{\'\i}ez et~al.}{2020}]{diez:2020}
D{\'\i}ez, V.~E., M.~Maier, J.~A. M{\'e}ndez-Zavaleta, and M.~T. Tehrani
  (2020).
\newblock Lagrangian constraint analysis of first-order classical field
  theories with an application to gravity.
\newblock {\em Physical Review D\/}~{\em 102\/}(6), 065015.

\bibitem[\protect\citeauthoryear{Dirac}{Dirac}{1958}]{Dirac:1958a}
Dirac, P. A.~M. (1958).
\newblock Generalized hamiltonian dynamics.
\newblock {\em Proceedings of the Royal Society of London. Series A,
  Mathematical and Physical Sciences\/}~{\em 246}, 333--3343.

\bibitem[\protect\citeauthoryear{Dirac}{Dirac}{1964}]{Dirac:1964}
Dirac, P. A.~M. (1964).
\newblock {\em Lectures on quantum mechanics}.
\newblock Dover Publications.

\bibitem[\protect\citeauthoryear{Fasso}{Fasso}{2005}]{fasso:2005}
Fasso, F. (2005).
\newblock Superintegrable hamiltonian systems: geometry and perturbations.
\newblock {\em Acta Applicandae Mathematica\/}~{\em 87\/}(1-3), 93--121.

\bibitem[\protect\citeauthoryear{Feintzeig}{Feintzeig}{2025}]{feintzeig:2025}
Feintzeig, B. (2025).
\newblock Quantization and the preservation of structure across theory change.
\newblock {\em Philosophy of Science\/}~{\em 92\/}(2), 259--284.

\bibitem[\protect\citeauthoryear{Feintzeig}{Feintzeig}{2024}]{feintzeig:2024b}
Feintzeig, B.~H. (2024).
\newblock Quantization as a categorical equivalence.
\newblock {\em Letters in Mathematical Physics\/}~{\em 114\/}(1), 19.

\bibitem[\protect\citeauthoryear{Feintzeig and Steeger}{Feintzeig and
  Steeger}{2024}]{feintzeig:2024}
Feintzeig, B.~H. and J.~Steeger (2024).
\newblock Classical limits of hilbert bimodules as symplectic dual pairs.
\newblock {\em Reviews in Mathematical Physics\/}~{\em 36\/}(10), 2450026.

\bibitem[\protect\citeauthoryear{Gitman and Tyutin}{Gitman and
  Tyutin}{}]{gitman:1990}
Gitman, D. and I.~V. Tyutin.
\newblock {\em Quantization of fields with constraints}.
\newblock Springer.

\bibitem[\protect\citeauthoryear{Gryb}{Gryb}{2025}]{gryb:2025}
Gryb, S. (2025).
\newblock Gauge symmetry and the arrow of time: How to count what counts.
\newblock {\em arXiv preprint arXiv:2509.14720\/}.

\bibitem[\protect\citeauthoryear{Gryb and Th{\'e}bault}{Gryb and
  Th{\'e}bault}{2026a}]{Gryb:2026}
Gryb, S. and K.~P. Th{\'e}bault (2026a).
\newblock Initial value constraints and the dirac-noether correspondence: A
  geometric account.
\newblock {\em (unpublished)\/}.

\bibitem[\protect\citeauthoryear{Gryb and Th{\'e}bault}{Gryb and
  Th{\'e}bault}{2016}]{Gryb:2016a}
Gryb, S. and K.~P.~Y. Th{\'e}bault (2016).
\newblock {Schr{\"o}dinger Evolution for the Universe: Reparametrization}.
\newblock {\em Classical and Quantum Gravity\/}~{\em 33\/}(6), 065004.

\bibitem[\protect\citeauthoryear{Gryb and Th{\'e}bault}{Gryb and
  Th{\'e}bault}{2024}]{gryb:2024}
Gryb, S. and K.~P.~Y. Th{\'e}bault (2024).
\newblock {\em Time Regained: Volume 1: Symmetry and Evolution in Classical
  Mechanics}, Volume~1.
\newblock Oxford University Press.

\bibitem[\protect\citeauthoryear{Gryb and Th{\'e}bault}{Gryb and
  Th{\'e}bault}{2026b}]{Gryb:2026c}
Gryb, S.~B. and K.~P. Th{\'e}bault (2026b).
\newblock Against the frozen formalism.
\newblock {\em (unpublished)\/}.

\bibitem[\protect\citeauthoryear{Halvorson}{Halvorson}{2012}]{Halvorson2012}
Halvorson, H. (2012).
\newblock What scientific theories could not be.
\newblock {\em Philosophy of Science\/}~{\em 79\/}(2), 183--206.

\bibitem[\protect\citeauthoryear{Halvorson}{Halvorson}{2016}]{Halvorson2016}
Halvorson, H. (2016).
\newblock Scientific theories.
\newblock In P.~Humphreys (Ed.), {\em The Oxford Handbook of Philosophy of
  Science}. Oxford University Press.

\bibitem[\protect\citeauthoryear{Henneaux and Teitelboim}{Henneaux and
  Teitelboim}{1992}]{Henneaux:1992a}
Henneaux, M. and C.~Teitelboim (1992).
\newblock {\em Quantization of gauge systems}.
\newblock Princeton University Press.

\bibitem[\protect\citeauthoryear{Isham}{Isham}{1993}]{isham:1993}
Isham, C.~J. (1993).
\newblock Canonical quantum gravity and the problem of time.
\newblock In {\em Integrable systems, quantum groups, and quantum field
  theories}, pp.\  157--287. Springer.

\bibitem[\protect\citeauthoryear{Kosmann-Schwarzbach}{Kosmann-Schwarzbach}{2010}]{kosmann:2010}
Kosmann-Schwarzbach, Y. (2010).
\newblock {\em The Noether Theorems: Invariance and Conservation Laws in the
  Twentieth Century}.
\newblock Springer Science \& Business Media.

\bibitem[\protect\citeauthoryear{Kucha{\v{r}}}{Kucha{\v{r}}}{1991}]{Kuchar:1991}
Kucha{\v{r}}, K.~V. (1991).
\newblock The problem of time in canonical quantization of relativistic
  systems.
\newblock In A.~Ashtekar and J.~Stachel (Eds.), {\em Conceptual Problems of
  Quantum Gravity}, pp.\  141. Boston University Press.

\bibitem[\protect\citeauthoryear{Landsman}{Landsman}{2001}]{landsman:2001}
Landsman, N. (2001).
\newblock Quantized reduction as a tensor product.
\newblock In {\em Quantization of singular symplectic quotients}, pp.\
  137--180. Springer.

\bibitem[\protect\citeauthoryear{Landsman}{Landsman}{2005}]{landsman:2005}
Landsman, N. (2005).
\newblock Functorial quantization and the guillemin-sternberg conjecture.
\newblock {\em Twenty years of Bialowieza: a mathematical anthology\/}~{\em 8},
  23--45.

\bibitem[\protect\citeauthoryear{Lusanna}{Lusanna}{1990}]{lusanna:1990a}
Lusanna, L. (1990).
\newblock An enlarged phase space for finite-dimensional constrained systems,
  unifying their lagrangian, phase-and velocity-space descriptions.
\newblock {\em Physics Reports\/}~{\em 185\/}(1), 1--54.

\bibitem[\protect\citeauthoryear{Lusanna}{Lusanna}{1991}]{lusanna:1991}
Lusanna, L. (1991).
\newblock The second noether theorem as the basis of the theory of singular
  lagrangians and hamiltonians constraints.
\newblock {\em La Rivista del Nuovo Cimento (1978-1999)\/}~{\em 14\/}(3),
  1--75.

\bibitem[\protect\citeauthoryear{Nguyen, Teh, and Wells}{Nguyen
  et~al.}{2020}]{nguyen:2020}
Nguyen, J., N.~J. Teh, and L.~Wells (2020).
\newblock Why surplus structure is not superfluous.
\newblock {\em The British Journal for the Philosophy of Science\/}.

\bibitem[\protect\citeauthoryear{Olver}{Olver}{1991}]{Olver:2000}
Olver, P.~J. (1991).
\newblock {\em Applications of Lie groups to differential equations}, Volume
  107.
\newblock Springer Science \& Business Media.

\bibitem[\protect\citeauthoryear{Ortega and Ratiu}{Ortega and
  Ratiu}{2013}]{ortega:2013}
Ortega, J.-P. and T.~S. Ratiu (2013).
\newblock {\em Momentum maps and Hamiltonian reduction}, Volume 222.
\newblock Springer Science \& Business Media.

\bibitem[\protect\citeauthoryear{Pitts}{Pitts}{2014}]{Pitts2014b}
Pitts, J.~B. (2014).
\newblock A first class constraint generates not a gauge transformation, but a
  bad physical change: The case of electromagnetism.
\newblock {\em Annals of Physics\/}~{\em 351}, 382--406.

\bibitem[\protect\citeauthoryear{Pitts}{Pitts}{2022}]{Pitts2022}
Pitts, J.~B. (2022).
\newblock First-class constraints, gauge transformations, de-ockhamization, and
  triviality: Replies to critics, or, how (not) to get a gauge transformation
  from a second-class primary constraint.

\bibitem[\protect\citeauthoryear{Pitts}{Pitts}{2024}]{Pitts2024}
Pitts, J.~B. (2024).
\newblock Does a second-class primary constraint generate a gauge
  transformation? electromagnetisms and gravities, massless and massive.
\newblock {\em Annals of Physics\/}~{\em 462}, 169621.

\bibitem[\protect\citeauthoryear{Pons}{Pons}{2005}]{Pons:2005}
Pons, J. (2005).
\newblock On dirac's incomplete analysis of gauge transformations.
\newblock {\em Studies In History and Philosophy of Science Part B:
  {\ldots}\/}~{\em 36}, 491.

\bibitem[\protect\citeauthoryear{Pons, Salisbury, and Sundermeyer}{Pons
  et~al.}{2010}]{pons:2010}
Pons, J., D.~Salisbury, and K.~A. Sundermeyer (2010).
\newblock Observables in classical canonical gravity: folklore demystified.
\newblock {\em Journal of Physics A: Mathematical and General\/}~{\em
  222\/}(12018).

\bibitem[\protect\citeauthoryear{Pooley}{Pooley}{2017}]{pooley:2017}
Pooley, O. (2017).
\newblock Background independence, diffeomorphism invariance and the meaning of
  coordinates.
\newblock In {\em Towards a theory of spacetime theories}, pp.\  105--143.
  Springer.

\bibitem[\protect\citeauthoryear{Pooley and Wallace}{Pooley and
  Wallace}{2022}]{PooleyWallace2022}
Pooley, O. and D.~Wallace (2022).
\newblock First-class constraints generate gauge transformations in
  electromagnetism (reply to pitts).

\bibitem[\protect\citeauthoryear{Read}{Read}{2024}]{read:2024}
Read, J. (2024).
\newblock {\em Background independence in classical and quantum gravity}.
\newblock Oxford University Press.

\bibitem[\protect\citeauthoryear{Salisbury and Sundermeyer}{Salisbury and
  Sundermeyer}{2017}]{Salisbury:2017}
Salisbury, D. and K.~Sundermeyer (2017).
\newblock L{\'e}on rosenfeld's general theory of constrained hamiltonian
  dynamics.
\newblock {\em The European Physical Journal H\/}~{\em 42\/}(1), 23--61.

\bibitem[\protect\citeauthoryear{Sudarshan and Mukunda}{Sudarshan and
  Mukunda}{1974}]{sudarshan:1974}
Sudarshan, E. C.~G. and N.~Mukunda (1974).
\newblock {\em Classical dynamics: a modern perspective}.
\newblock World Scientific.

\bibitem[\protect\citeauthoryear{Sundermeyer}{Sundermeyer}{1982}]{sundermeyer:1982}
Sundermeyer, K. (1982).
\newblock {\em Constrained dynamics with applications to Yang-Mills theory,
  general relativity, classical spin, dual string model}.
\newblock Springer-Verlag.

\bibitem[\protect\citeauthoryear{Th{\'e}bault}{Th{\'e}bault}{2011}]{Thebault:2011b}
Th{\'e}bault, K. P.~Y. (2011, May).
\newblock Symplectic reduction and the problem of time in nonrelativistic
  mechanics.

\bibitem[\protect\citeauthoryear{Th\'{e}bault}{Th\'{e}bault}{2026}]{Thebault2026}
Th\'{e}bault, K. P.~Y. (2026).
\newblock {\em Classical and Quantum Phase Space Mechanics}.
\newblock Elements in the Philosophy of Physics. Cambridge: Cambridge
  University Press.

\bibitem[\protect\citeauthoryear{Wallace}{Wallace}{2019}]{Wallace:2019}
Wallace, D. (2019).
\newblock Observability, redundancy and modality for dynamical symmetry
  transformations.

\bibitem[\protect\citeauthoryear{Weatherall}{Weatherall}{2018}]{weatherall:2018}
Weatherall, J.~O. (2018).
\newblock Regarding the `hole argument'.
\newblock {\em The British Journal for the Philosophy of Science\/}~{\em
  69\/}(2), 329--350.

\bibitem[\protect\citeauthoryear{Weatherall}{Weatherall}{2019a}]{Weatherall2019a}
Weatherall, J.~O. (2019a).
\newblock Theoretical equivalence in physics: Part 1.
\newblock {\em Philosophy Compass\/}~{\em 14\/}(5), e12592.

\bibitem[\protect\citeauthoryear{Weatherall}{Weatherall}{2019b}]{Weatherall2019b}
Weatherall, J.~O. (2019b).
\newblock Theoretical equivalence in physics: Part 2.
\newblock {\em Philosophy Compass\/}~{\em 14\/}(5), e12591.

\bibitem[\protect\citeauthoryear{Weinstein}{Weinstein}{1983}]{weinstein:1983}
Weinstein, A. (1983).
\newblock The local structure of poisson manifolds.
\newblock {\em Journal of differential geometry\/}~{\em 18\/}(3), 523--557.

\bibitem[\protect\citeauthoryear{Woodhouse}{Woodhouse}{1997}]{woodhouse:1997}
Woodhouse, N. M.~J. (1997).
\newblock {\em Geometric quantization}.
\newblock Oxford University Press.

\end{thebibliography}

\end{document}